\documentclass[aps,prd,twocolumn]{revtex4}
\usepackage{epsfig}

\usepackage{amssymb}
\usepackage{wasysym}  
\usepackage{dsfont} 

\providecommand{\jup}{\parbox{1ex}{{\scalebox{0.6}{\jupiter\hspace*{2pt}}}}}
\providecommand{\ven}{\parbox{1ex}{{\scalebox{0.6}{\venus\hspace*{2pt}}}}}
\providecommand{\ert}{\parbox{1ex}{{\scalebox{0.6}{$\oplus$\hspace*{2pt}}}}}

\providecommand{\vect}[1]{\mathbf{#1}}

\begin{document}

\title{WIMP diffusion in the solar system including solar depletion and its effect on Earth capture rates}
\author{Johan Lundberg}
\email{johan@physto.se}
\affiliation{Division of High Energy Physics, Uppsala University, SE-751 21 Uppsala, Sweden}

\author{Joakim Edsj\"o}
\email{edsjo@physto.se}
\affiliation{Department of Physics, AlbaNova University Center,
Stockholm University, SE-106 91 Stockholm, Sweden}

\date{\today}

\begin{abstract}

Weakly Interacting Massive Particles (WIMPs) can be captured by the
Earth, where they eventually sink to the core, annihilate and produce
e.g.\ neutrinos that can be searched for with neutrino telescopes. The
Earth is believed to capture WIMPs not dominantly from the Milky Way
halo directly, but instead from a distribution of WIMPs that have
diffused around in the solar system due to gravitational interactions
with the planets in the solar system. Recently, doubts have been
raised about the lifetime of these WIMP orbits due to solar capture.
We here investigate this issue by detailed numerical simulations.

Compared to earlier estimates, we find that the WIMP velocity distribution 
is significantly suppressed below about
70 km/s which results in a suppression of the capture rates mainly 
for heavier WIMPs (above $\sim$ 100 GeV). At 1 TeV and above the reduction
is almost a factor of 10.

We apply these results to the case where the WIMP is a supersymmetric
neutralino and find that, within the Minimal Supersymmetric Standard
Model (MSSM), the annihilation rates, and thus the neutrino fluxes, 
are reduced even more than the capture rates.
At high masses (above $\sim$ 1 TeV), the suppression is almost two orders of magnitude.

This suppression will make the detection of neutrinos from heavy WIMP annihilations in the Earth much harder compared to earlier estimates.

\end{abstract}


\maketitle

\section{Introduction}

There is mounting evidence that a major fraction of the matter in the Universe is dark. The WMAP Experiment gives as a best fit value that \cite{wmap} $\Omega_{\rm CDM} h^2=0.113\pm0.009$, where $\Omega_{\rm CDM}$ is the relic density of cold dark matter in units of the critical density and $h$ is the Hubble parameter in units of 100 km\,s$^{-1}$\,Mpc$^{-1}$. One of the main candidates for the dark matter is a Weakly Interacting Massive Particle (WIMP), of which the supersymmetric neutralino is a favorite candidate. 
There are many ongoing efforts trying to find these dark matter particles, either via direct detection or via indirect detection by detecting their annihilation products.

One of the proposed search strategies is to search for a flux of high-energy neutrinos from the center of the Earth \cite{earth-nus}. This idea goes back to Press and Spergel \cite{press-spergel}, who calculated the capture rate of heavy particles by the Sun.
For the Earth, the idea is that WIMPs can scatter off a nucleus in the Earth, lose enough energy to be gravitationally trapped, eventually sink to the core due to subsequent scatters, annihilate and produce neutrinos. Due to purely kinematical reasons, the capture rate in the Earth depends strongly on the mass and the velocity distribution of the WIMPs. The heavier the WIMP is, the lower the velocity needs to be to facilitate capture. In Ref.\ \cite{gould-resonant}, Gould refined the calculations of Press and Spergel for the Earth and derived exact formulae for the capture rates. In a later paper \cite{gould-direct}, Gould pointed out that since the Earth is in the gravitational potential of the Sun, all WIMPs will have gained velocity when they reach the Earth and hence capture of heavy WIMPs would be very small. However, Gould later realized \cite{gould-diff} that due to gravitational interactions with the other planets (mainly Jupiter, Venus and Earth), WIMPs will diffuse in the solar system both between different bound orbits, but also between unbound and bound orbits. Gould showed that the net result of this is that the velocity distribution at the Earth will effectively be the same as if the Earth was in free space. This approximation is widely used today where one further assumes that the halo velocity distribution is Gaussian (i.e.\ a Maxwell-Boltzmann distribution).

However, Farinella et al.\ \cite{farinella} later made simulations of Near Earth Asteroids (NEAs) that had been ejected from the asteroid belt. They found that many of these have life times of less than two million years. After that time they are either thrown into the sun or thrown out of the solar system. If this typical lifetime also applies to WIMPs, this would significantly reduce the number of WIMPs bound in the solar system, as pointed out in Ref.\ \cite{gould-conserv}. This in turn would reduce the expected capture and annihilation rates in the Earth and thus reduce the neutrino fluxes. In Ref.~\cite{gould-conserv}, Gould and Alam investigated what the implications would be if bound WIMPs would actually be thrown into Sun. They investigated two scenarios: an \emph{ultra conservative} scenario where all bound WIMPs are depleted and a \emph{conservative} scenario where all bound WIMPs that do not have Jupiter-crossing orbits are depleted. In the ultra conservative view solar depletion is assumed to be so efficient that no bound WIMPs exist, whereas in the conservative view, Jupiter is assumed to be faster at diffusing WIMPs into the solar system than solar depletion is at throwing them into the Sun. Both of these views significantly reduce the neutrino fluxes from the Earth for heavier WIMPs. 

However, the truth probably lies somewhere between the conservative view and the assumption that solar depletion is very inefficient, i.e.\ some WIMPs on bound orbits in the inner solar system will survive, but solar capture will diminish their numbers somewhat. The aim of this
paper is to investigate the effects of solar capture on the distribution of WIMPs in the solar system and the implication this has on expected neutrino fluxes from the Earth. We will do this by numerical simulations of WIMPs in the solar system and by reanalyzing the process of WIMP diffusion in the solar system. Finally, we will apply our results to the case where the WIMP is the neutralino, which arises naturally in Minimal Supersymmetric extensions of the Standard Model (MSSM).

The layout of this paper is as follows. In section \ref{sec:capintro}, we will briefly review the history of WIMP capture calculations for the Earth. In section \ref{sec:galhalodiff} we will go through our assumed halo model and the role of diffusion in more detail. In section \ref{sec:one-diff} we will go through the formalism for the diffusion caused by one planet and
in section \ref{sec:soldepletion} we add the new ingredient, solar depletion. In 
section \ref{sec:nummethods} we present our numerical treatment of the diffusion problem. All of this will be put together with the dominant planets for diffusion in section 
\ref{sec:combined-diff} where our main results on the velocity distribution at the Earth are presented. In the remaining sections we will investigate how this affects the capture and annihilation rates in the Earth and will present results on the expected neutrino-induced muon fluxes in MSSM models in section \ref{sec:susy}. Finally, we will conclude in section \ref{sec:conclusions}.


\section{Capture of WIMPs by the Earth -- historical remarks}
\label{sec:capintro}

Capture of WIMPs by the Sun was first studied by Press and Spergel 1985 \cite{press-spergel}. Their calculations were approximate in nature, especially when applied to the Earth. This was refined in a series of papers by Gould \cite{gould-resonant,gould-direct,gould-diff}. 
In 1987, Gould \cite{gould-resonant} derived the exact formulae needed to calculate
the capture of WIMPs by a spherically symmetric body. When applied to
the capture by the Sun and the Earth, his approach enhanced the capture by
factors of 1.5--3 and 10--300 respectively, compared to the previous
approximations by Press and Spergel \cite{press-spergel}.

However, in 1988, Gould \cite{gould-direct} refined the analysis, taking into account
that the Earth is well inside the gravitational potential of the Sun.
The velocities of the incoming particles are increased when they
approach the potential of the Sun. This reduces capture substantially.
On the other hand, bound solar orbits are allowed. Gould realized that 
particles scattered by the Earth could become bound to the solar system. 
This scattering can be of two kinds: \emph{gravitational scattering}, which is elastic in that the velocity with respect to the Earth is conserved or \emph{weak scattering} off an atom, which can be inelastic and either lead to capture by the Earth or make the particle bound to the solar system. In this context, an equation for estimating the timescales of weak and gravitational scattering was developed, following traditions of \"Opik \cite{opik}.

Among other things, Gould concluded that, due to the differences in total
scattering cross section, the gravity of the Earth is more effective
in changing the orbits of bound particles than is weak scattering. For capture by the Earth though, weak scattering is the only process at work since gravitational scattering leaves the velocity with respect to the Earth unchanged. 

In 1991, Gould \cite{gould-diff} continued further, moving his attention to the
gravitational diffusion caused by the other planets. Further, he considered
the combined diffusion effect of Jupiter, Venus and the Earth concluding that it
will make the velocity distribution isotropic in the frame of the
Earth. Based essentially on Liouville's theorem, this means that the phase
space density of unbound and bound
particles would be the same. Specifically, for the most important
parts of velocity space, this would happen on time scales shorter than
the age of the Solar System. Obviously, such a scenario would
substantially enhance capture of heavy WIMPs by the Earth. Further,
he concluded that weak capture of WIMPs to bound solar orbits is
negligible, and that one may use the ''free-space'' formulae derived in Ref.\ \cite{gould-resonant} for capture, even though the Earth is deep within the potential well of
the Sun. 

As mentioned in the introduction, the calculations took a new unexpected turn in 1999 when Gould and Alam interpreted \cite{gould-conserv} the results of Farinella et al.~\cite{farinella}.
Farinella et al.\ had
numerically calculated the fates of about 50 asteroids of which most
were considered to be \emph{near Earth asteroids} (NEAs). They
concluded that about a third of the considered asteroids will be
ejected to hyperbolic orbits or, more importantly, driven into the sun
in less than 2 million years. If the results of Farinella et al.\ were
applicable to general Earth crossing orbits of WIMPs, the part of
velocity space corresponding to bound solar orbits would be
effectively empty, since the typical time scales at which such orbits are
populated from unbound orbits are generally much longer \cite{gould-diff}. The basic
results of Farinella et al.\ were later confirmed by Gladman et al.\ \cite{gladman} 
and Migliorini et al.\ \cite{migliorini}.

To investigate the role of Solar depletion, Gould and Alam \cite{gould-conserv} analytically
investigated the difference between the 1991 case of no solar
depletion, and the other extreme, where there is no dark matter in
solar system bound orbits at the Earth. In the latter
\emph{ultra conservative view}, capture is
heavily suppressed. For instance, they found that WIMPs with masses above about 325 GeV
could not be captured by the Earth at all. They also considered a scenario
where WIMPs in Jupiter--crossing orbits were not effected by solar
capture, the \emph{conservative view}, which they found allows WIMPs up to about 630 GeV
to be captured. (See section \ref{sec:gravdiff} below for a more detailed discussion of these cutoff masses.) Both of these scenarios significantly suppress the capture rates of WIMPs by the Earth and the question to ask is if the results of Farinella et al.\ can really be applied to all Earth--crossing WIMP orbits? The orbits of asteroids ejected from the asteroid belt are, after all, rather special as they typically arise from resonances. It is thus not necessarily so that these results apply to all bound WIMPs. We will in the coming sections go through the necessary steps to investigate this question in detail.

\section{The galactic halo model and cutoff masses}
\label{sec:galhalodiff}

\subsection{The galactic halo model}
\label{sec:galhalo}

In order to make the calculations concrete, we use the Maxwell-Boltzmann
model \cite{susydm},
where the local velocity distribution of WIMPs is Gaussian in the
inertial frame of the Galaxy. At the location of the Sun the distribution is 
\begin{equation}
  f_v(v) \textrm{d}^3v = \frac{e^{-v^2/v_0}}{\pi^{3/2}v_0^3} \textrm{d}^3v, 
\end{equation}
where $v_0=\sqrt{\frac{2}{3}}\bar{v}$ with $\bar{v}$ being the three-dimensional velocity dispersion.
We will here use the standard value of $\bar{v}=270$ km/s corresponding to $v_0= 220$ km/s.
The distribution is normalized such that 
\begin{equation}
\int f_v(v) 4\pi v^2 \textrm{d}v =1.
\end{equation}

The velocity distribution can be galileo transformed into
the frame of the Sun: $f_s(s)$, where
$\vect{s}=\vect{v}+\vect{v}_\textrm{\footnotesize{}Sun}$, and
$v_\textrm{\footnotesize{}Sun}$=220 km/s, and averaged over
all angles. In this special case of a Gaussian distribution the
transformation can be done in closed form\cite{gould-resonant}. As
Gould have pointed out, the angle between the rotation axis of the
solar system and that of the galaxy is about 60$^\circ$ which makes
the velocity distribution very close to spherically symmetric,
\emph{if one considers averages over a galactic year} $\approx$ 200
million years\cite{gould-direct}. The distribution used is mirror
symmetric in the galactic plane which means that the time of average
need only be 100 million years.%

The symbol $F_s(s)$ will be used to denote the phase space number
density
\begin{equation}
  F_s(s) = \frac{\rho_\chi}{M_\chi} f_s(s),
\label{solgauss}
\end{equation}
where $M_\chi$ is the WIMP mass, and $\rho_\chi$ is the WIMP
mass per unit volume in the halo.
When the particles of this distribution pass through the solar
system, the velocities are boosted and focused by to the gravitational potential.
At the location of the Earth, the solar system escape velocity
is $\sqrt{2}v_{{\ert}}\approx$42 km/s, where we have used the speed of the Earth, $v_{\ert} \simeq 29.8$ km/s. Therefore the velocity at the location of the Earth, $w$, is, according
to conservation of energy
\begin{equation}
  w^2 = s^2 + 2 v_{{\ert}}^2.
\end{equation}

When a spherically
symmetric distribution such as $F_s(s)$ is focused by a Coulomb
potential such as that of the Sun, the following statement holds:
\cite{gould-direct}
\begin{equation}
\label{focusing}
  \frac{ F_w(w) 4\pi w^2 dw}{w} = \frac{ F_s(s) 4\pi s^2 ds}{s}
\end{equation}
This can be understood as Liouville's theorem for
the spherically averaged phase space density, since
\begin{equation}
\frac{ds}{dw}=\frac{w}{s} \Rightarrow F_w(w) =  F_s(s).
\end{equation}
Since the velocity $w$ of the halo particles is always at least equal
to the escape velocity, there will be a hole in velocity space
so that
\begin{equation}
 F_w(w) =0 \textrm{ when } w< \sqrt{2}v_{{\ert}}.
\end{equation}
This is important since capture by the
Earth is very sensitive to $F_w(w)$ at low velocities.

\begin{figure}
\center
\epsfig{width=\columnwidth,file=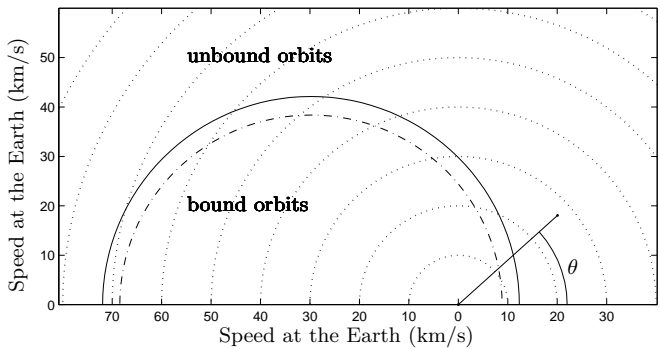}
\caption{The ecliptic ($\phi=\pi/2$) slice of the particle velocity space
  \emph{in the frame of the Earth}. The dotted curves show the velocity relative to
  the Earth and the indicated angle $\theta$, is the angle of the particle with respect to
  the direction of Earth's motion. The angle $\phi$ determines in which angle we cut the
  velocity sphere. $\phi=0$ is the south pole of the solar system and $\phi=\pi/2$ (as 
  shown here) is the slice radially outward from the Earth.
  The region inside the solid
  semicircle represents bound orbits. It's radius is the escape
  velocity from the Solar system at the location of the Earth, but in
  the frame of the Sun. In the same way, the region outside the
  dash-dotted line (an almost perfect semi-circle) corresponds to particles that
  may reach Jupiter. By
  repeated close encounters with the Earth, particles may diffuse
  along the dotted circles (actually spheres) of constant velocity
  only, keeping $u$ constant, but allowing changes in $\theta$ and
  $\phi$, as explained in the text. \label{pic_diffprincip}}
\end{figure}

The distribution $F_w(w)$ can now be used to calculate the
distribution as seen from the moving Earth where the particle velocity
is $\vect{u}=\vect{w}+\vect{v}_{{\ert}}$
\begin{equation}
  F_u(\vect{u})=F_w(\vect{w})=F_w(\vect{u-v_{{\ert}}}).
\label{used_freedens}
\end{equation}
This means that the \emph{hole} is shifted, so that it is centered
around $-\vect{v}_{{\ert}}$. This is visualized by figure
\ref{pic_diffprincip} which displays a two dimensional slice
of the three dimensional velocity space.

\subsection{Cutoff masses when low velocity WIMPs are missing}
\label{sec:gravdiff}

\begin{figure}
\center
\epsfig{width=\columnwidth,file=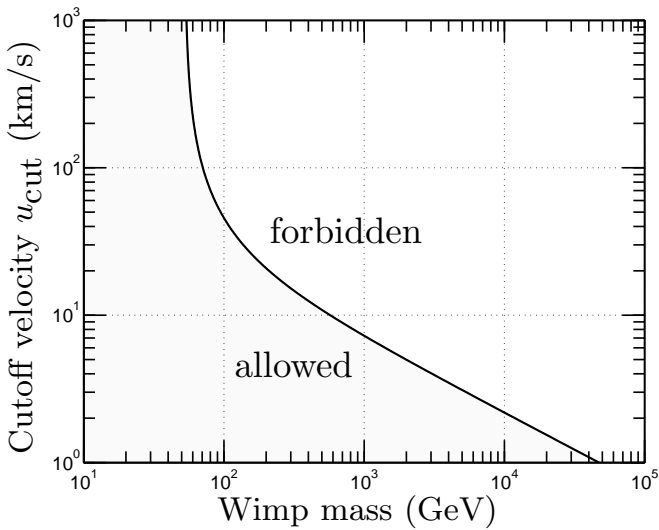}
\caption{Cutoff velocity $v_\textrm{cut}$ against WIMP mass $M$. Only
combinations of $M$ and $v_\textrm{cut}$ to the left of the line
is kinematically allowed (in the sense that they can lead to capture by the Earth).
 \label{vofmplot}}
\end{figure}

In the absence of WIMPs gravitationally bound to the solar system, the
capture by the Earth is totally suppressed for WIMP masses larger
than a critical value. To understand this, consider a particle
approaching the Earth with velocity $u$ at infinity with respect to
the gravitational potential of the Earth. If it is to be captured, it
must be scattered by an atom to a velocity less than the escape
velocity $v_\textrm{esc}$ at the atom. By conservation of energy and
momentum, the particle must have a velocity less than (assuming iron
to be the heaviest relevant element of the Earth)
\begin{equation}
  u_\textrm{cut} = 2 \frac{\sqrt{M_\chi M_\textrm{\small Fe}}}%
{M_\chi-M_\textrm{\small Fe}} v_\textrm{esc}\label{vcut}
\end{equation}
when it approaches the gravitational potential of the Earth. 
Solving for the WIMP mass $M_\chi$ gives
\begin{equation}
  M_{\chi\textrm{,cut}} = M_\textrm{\small Fe}\frac{ u^2 + 2
  v_\textrm{esc} \ ( v_\textrm{esc} + \sqrt{u^2 +
  v_\textrm{esc}^2 } )}{ u^2}.\label{mcutofu}
\end{equation}
Here, $u$ is the speed (at infinity) of the approaching particle in the frame of the
Earth and \mbox{$M_{\chi\textrm{,cut}}$} is the highest allowed
mass of the particle if it is to be captured by the Earth. The escape velocity
varies from 11.2 km/s at the surface to 15.0 km/s at the center of the Earth (see section \ref{sec:newcap} for more information about the Earth model we use), and
capture is thus easiest at the center where the escape velocity is higher.
Using $v_{\rm esc}=15.0$ km/s, we plot in Fig.~\ref{vofmplot} the relation between $u_{\rm cut}$ and the cutoff mass, $M_{\chi,cut}$.

With Eq.~(\ref{mcutofu}), we can now relate to the cutoff masses in the conservative and ultra conservative views by Gould and Alam \cite{gould-conserv}.
In the ultra conservative view, we assume that only unbound halo particles are captured.
Halo particles cannot be slower than $u_{\rm cut} = (\sqrt{2}-1)v_{\ert}\simeq 12.3$ km/s at, and in the frame of, the Earth (this is also seen in Fig.~\ref{pic_diffprincip}).
This gives a cutoff mass of about $410$ GeV over which
capture by the Earth is impossible. This differs from the value of 325 GeV for the ultra conservative view in Gould and Alam \cite{gould-conserv}. The difference is because they used an average escape velocity of 13 km/s instead of the maximal one of 15 km/s that we have used in Fig.~\ref{vofmplot}. 

In the conservative view, we assume that Jupiter--crossing orbits are filled. This means that all orbits outside the dot-dashed curve and the dashed curve in Fig.~\ref{pic_diffprincip} are filled. The lowest velocity WIMP at the Earth that is on a Jupiter--crossing orbit is in the lower right-hand end of the dot-dashed curve and it has a velocity of $u_{\rm cut} = v_{\ert} (\sqrt{2/(1-r_{\ert}/r_{\jup})} -1) \simeq 8.8$ km/s (and is moving in the same direction as the Earth). $r_{\jup} \simeq 5.2 r_{\ert}$ is the radius of the Jupiter orbit. This value of $u_{\rm cut}$ gives a cutoff mass of about 712 GeV, whereas Gould and Alam \cite{gould-conserv} got a cutoff mass of about 630 GeV. The difference is again due to the different escape velocities used, but also a different velocity to reach Jupiter. We use the value $u_{\rm cut}\simeq 8.8$ km/s as indicated above, whereas they used an approximation for more general orbits than the one giving the cutoff derived here. So, to conclude, in the conservative and ultra conservative view, we cannot capture WIMPs heavier than about 410 GeV and 712 GeV respectively. This is in rough agreement with the results of Gould and Alam \cite{gould-conserv}.

If, on the other hand, the solar
system is full of gravitationally bound dark matter, the velocities
can be much lower. As the lowest allowed velocity of the WIMPs
$u_\textrm{cut}$ tends to zero, the mass limit
\mbox{$M_{\chi\textrm{,cut}}$} goes to infinity. Typically,
most WIMPs in the Galaxy have velocities much greater than those of
Eq.~(\ref{vcut}), so only a small fraction of the WIMPs are possible to capture.

A particle in close encounter with a planet, for instance the Earth,
may get gravitationally scattered into a new direction and a new
velocity \emph{as seen from the frame of the Sun}. However, by
conservation of energy, the speed $u$ \emph{with respect to the frame
of the planet} is unchanged. This means that a particle at a particular
place in velocity space may, by repeated close encounters with the
Earth, diffuse to any location on the sphere of constant velocity (with respect to the Earth), and nowhere else.

The location of a particle at such a sphere can be specified by the
angles at which it passes the Earth. The angle $\theta$ is
measured between the forward direction of the Earth and the velocity
vector of the particle, and $\phi$ is the angle of rotation
around the forward direction of the Earth, with $\phi=0$ at the North
pole of the solar system.

Figure \ref{pic_diffprincip} illustrates how spheres (and circles) of
constant $u$ cross the limit of where particles have bound and unbound
orbits. This corresponds to the possibility of gravitational capture
and ejection from the solar system.

A single planet can diffuse particles along spheres of constant
velocity only. Therefore, it is clear (from e.g.\ Fig.~\ref{pic_diffprincip})
that orbits with 
velocities $u$ less than $12.3$ km/s cannot be populated by the Earth
alone. However, since the velocity spheres of different planet's are
not concentric (they need not even be spheres when the particles reach
another planet), the combined effect may diffuse particles down to
Earth-crossing velocities $u$ less than $12.3$ km/s. This will be
investigated in detail in section
\ref{sec:combined-diff}. In order to do so, we must first understand
how a single planet affects the phase space distribution. 

\section{Gravitational diffusion in the one planet case}
\label{sec:one-diff}

In this section, we investigate the details of what will be called
gravitational diffusion. We will develop tools for detailed
investigation of the bound orbit phase space density, taking the
effects of solar depletion into account. We will here start by looking at
diffusion effects from a single planet only and will take the Earth as an example. The exact same formalism is then used for Venus and Jupiter as well.

In this section we assume that when a
particle is in Earth crossing orbit (perihelion less than the Earth
orbit radius $R_{\ert}$ and aphelion greater than $R_{\ert}$), long range
interactions with other planets are less important,
and can be ignored. This is not a problem, as we in section \ref{sec:combined-diff} add the effects of other planets (apart from possible resonances). 
We will in this section closely follow Gould \cite{gould-diff}, with some small modifications.

\subsection{The probability of planet collisions.}

We are interested in calculating the rate at which WIMPs with Earth
crossing orbits comes into close encounter with the Earth. This will
be used to estimate how the Earth affects the WIMP distribution. A
close encounter is an event were the particle's impact parameter is
smaller than or equal to some value $b_{max}(u)$.

\begin{figure}
\center
\epsfig{file=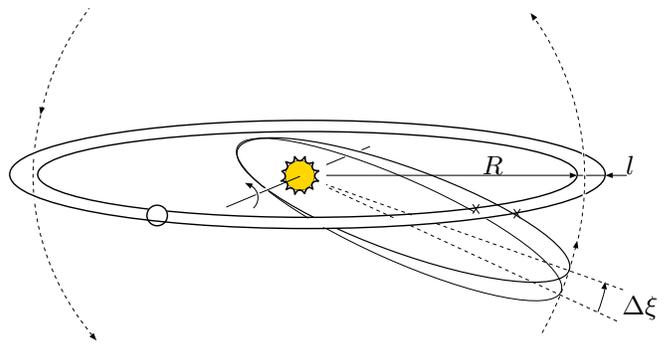, width=\columnwidth}
\caption{{The angle of perihelion precession $\Delta\xi$, as the orbit
  enters and leaves the disk of the Earth.} In this example, the plane
  of the orbit is nearly perpendicular to the ecliptic plane.
\label{pic_preiorb}}
\end{figure}
\noindent

Let's imagine the Earth as being spread out on a flat ring of inner
radius $R$, outer radius $R+l$ and thickness $h$, as in figure
\ref{pic_preiorb}. Now consider a particle with perihelion less than the
planet orbit radius $R$ and aphelion greater than $R$. Such particles
will be said to have Earth crossing orbits. This is motivated by the
fact that due to the precession of perihelion, all such orbit ellipses
will eventually intersect the Earth ring. The small angle the
perihelion sweeps out, as the orbit ellipse enters and leaves the ring is
given by
\begin{equation}
    \Delta\xi \approx \tan \Delta\xi = \frac{l}{R} | \tan \Theta_1 | 
\end{equation} 
where $\Theta_1$ is the intersection angle between the WIMP ellipse
and the plane perpendicular to the location vector of the Earth
$\vect{R}$. Since this happens four times during each perihelion
revolution, the mean probability for such a WIMP to intersect the ring of
the Earth during each WIMP year $T_\chi$ is
\begin{equation}
\left\langle p_{T_\chi} \right\rangle = \frac{4 \Delta\xi}{2\pi}.
\end{equation}
The probability for the WIMP to 
come into close encounter with the Earth is therefore $p_{T_\chi}$ times the cross section $\sigma$ of such an
event, divided by the area over which the Earth is distributed.
However, the
length of the path which is inside the Earth ring during each
encounter is $\propto |\cos \Theta_2|^{-1}$, where $\Theta_2$ is the
angle between the axis of the ecliptic and $\vect{u}$, the velocity of the
WIMP as seen from Earth \cite{gould-direct}. The probability for a reaction with
cross--section $\sigma$, can now be calculated,%
\begin{equation}
\frac{p(\sigma)}{T_{\ert}} =  \frac{1}{|\cos \Theta_2|} \frac{\sigma}{2\pi R l} \frac{4 l}{2\pi R} | \tan
\Theta_1 | \frac{1}{T_\chi}.
\end{equation}
where we have divided by $T_{\ert}$ to get the probability per unit time.
The WIMP year can be written in terms of
$\vect{u}(\theta,\phi,u)\,$\cite{gould-direct}, the velocity of the
particle in the frame of the Earth,
\begin{equation}
T_\chi=\big(1-2\frac{u}{v_{\ert}}\cos\theta-\frac{u^2}{v^2_{\ert}}\big)^{-3/2} T_{\ert},
\end{equation}
and $\Theta_1(\vect{u})$ and $\Theta_2(\vect{u})$ can be expressed in
$u$, $\theta$ and $\phi$:
\begin{eqnarray}
\cos\Theta_1 & = & \vect{\hat{R}}\cdot\vect{\hat{v}}_{\chi} = \frac{\vect{R} \cdot (\vect{u} +
\vect{v_{\ert}})}{R|\vect{u} + \vect{v_{\ert}}|} \nonumber \\
& =&  \frac{u \sin\theta\sin\phi}%
{(u^2+v^2_{\ert} + 2uv_{\ert} \cos\theta)^{1/2}},
\end{eqnarray}
\begin{equation}
\cos\Theta_2 = \sin\theta\cos\phi = \frac{\vect{u} \cdot (\vect{v}_{\ert}\vect{\times}\vect{R} ) }%
{uv_{\ert} R} ,
\end{equation}
\begin{eqnarray}
\lefteqn{\cot\Theta_1  =  \frac{u}{v_{\ert}}}\nonumber \\
& & \frac{\sin\theta\sin\phi}%
{\big(1+2(u/v_{\ert})\cos\theta+(u^2/v_{\ert}^2)(1-\sin^2\theta\sin^2\phi)\big)^{1/2}}.
\end{eqnarray}
By substituting and rearranging we conclude that the yearly reaction
probability for an event with cross--section $\sigma$ is given by
\begin{equation}
\label{genyearprob}
\frac{p(\sigma,\vect{u})}{T_{\ert}} = \frac{3}{2}\frac{ \sigma }{\pi
R^2}\frac{v_{\ert}}{u} \gamma(\vect{u})^{-1}, \textrm{where}
\end{equation}
\begin{eqnarray}
\lefteqn{\gamma(\vect{u})=} \nonumber \\
& & \frac{3}{2}\frac{\pi\sin^2\theta \: |\sin\phi\cos\phi| \: (1-2u/v_{\ert}\cos\theta-u^2/v^2_{\ert})^{-3/2}
}%
{\big(1+2(u/v_{\ert})\cos\theta+(u^2/v_{\ert}^2)(1-\sin^2\theta\sin^2\phi)\big)^{1/2}}.
\label{gammaekv}
\end{eqnarray}
Eq.~(\ref{gammaekv}) was first derived by Gould
\cite{gould-direct} in a very similar way. Among other things, he used
it to calculate the ''typical timescales'' at which particles diffuse
between different velocity space regions in the absence of solar
depletion. It is also used for calculating the probability of weak
scattering of WIMPs at the Earth.

The equations above are derived under some (geometrical) approximations with the aim of getting the correct scattering probabilities on average. There are however a few pathological cases where the geometrical model used above breaks down. This happens when $\phi=0$, $\phi=\pi/2$, $\theta=0$ and
$\theta=\pi$, in which case the probabilities above are unphysical. Since this only happens for these few special cases we will artificially solve this by adding a small angle (of about 1 degree) to $\theta$ and $\phi$ when close to these regions. Note that in principal, the problems could be resolved, by making
$\Delta\xi$ a function of the full set of orbit parameters, but this is unnecessarily complicated for our purposes. For the interested reader, we refer to a detailed investigation
of the mathematical properties of $\gamma$ as presented in
\cite{gould-direct} and \cite{gould-diff}. To test our solution of adding a small angle in these pathological cases, we have investigated the
effect of further increasing the small angle added and conclude that the actual value chosen is not important for the final results. This is reasonable, since orbits
in the vicinity of these critical regions are quickly deflected into other orbits anyway.


\subsection{Gravitational scattering on a planet}

Now that we have learnt how to calculate the probability for particles
to come into close encounter with a given planet, it is time to apply
this to gravitational diffusion. For the Earth, we were mainly
interested in those particles crossing the sphere of one AU during
each revolution, since they have a chance of hitting the Earth (and
possibly be weakly captured to it) within each perihelion precession
revolution.

The gravitational scattering probability is dependent on the angular
distance between the velocities before and after scattering:
$\vect{u}$ and $\vect{u}'$, such that small deflections are more
common. The angle can be related to the impact parameter $b$,
\begin{equation}
\delta(b)=\pi- 2 \arctan\frac{b u^2}{MG},
\label{deltaofb}
\end{equation}
as well as
\begin{equation}
\delta(\vect{\hat{u}',\hat{u}}) = \arccos(\vect{\hat{u}}'\cdot
\vect{\hat{u}}).
\end{equation}
where $\vect{\hat{~}}$ denotes unit vectors.
The scattering angle above is the one given by Rutherford scattering
(see e.g.\ \cite{scheck}). Gould used an approximate formula when deriving the typical
time scales \cite{gould-resonant}: $\delta(b)=R_{\oplus} v_{esc}^2 / (b u^2)$.
The two differ at very small impact parameters, and we use the full expression in our
calculations.

\begin{figure}
\center
\epsfig{width=\columnwidth,file=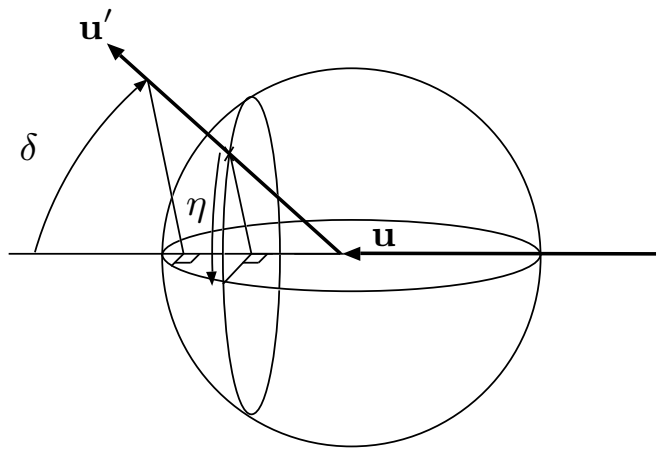}
\caption{{Scattering off the Earth in velocity space.} A fixed impact
  parameter fixes $\delta$, but $\eta$ is evenly distributed.
\label{pic_rudscat}}
\end{figure}
As mentioned before, scattering can only change the direction and not the velocity,
and we are therefore dealing with random walk on spheres
of constant $u$. The direction $\eta$ of the scattering is evenly
distributed, as seen in Fig.~\ref{pic_rudscat}, where the scattering setup is shown. 
The arc length
is fixed by $\delta(b)$, but the scattering \emph{direction} is
evenly distributed.

The cross--section for scattering between $\delta$ and $\delta +
d\delta$ is $d\sigma = 2\pi b db$, so the yearly probability for
scattering in this range is, (using Eq.~\ref{genyearprob})
\begin{equation}
\label{gravprobb}
\frac{dp(\vect{u},b)}{db \: T_{\ert}} = %
\frac{3}{2}\frac{ 2\pi b }{\pi
R^2} \frac{v_{\ert}}{u}\gamma(\vect{u})^{-1}.
\end{equation}
This can be rewritten in terms of the scattering angle $\delta(\vect{\hat{u}',\hat{u}})$,
\begin{eqnarray}
\lefteqn{\frac{dp(\vect{u},b(\delta))}{d\delta \: T_{\ert}} =} \nonumber \\%
& & - \frac{3}{2}\frac{ 2\pi }{\pi R^2} \frac{v_{\ert}}{u}\gamma(\vect{u})^{-1} %
\frac{(M_{\ert} G)^2}{u^4}
\frac{\phantom{2}\cos\phantom{^3}({\delta(\vect{\hat{u}',\hat{u}})}/{2})}{2
\sin^3({\delta(\vect{\hat{u}',\hat{u}})}/{2})}
\label{gravprobd}
\end{eqnarray}
since 
\begin{equation}
b=\frac{M_{\ert} G}{u^2}\cot\frac{\delta}{2} %
\quad ; \quad%
\frac{db}{d\delta}=-\frac{M_{\ert} G}{u^2} \frac{1}{2 \sin^2 \frac{\delta}{2} }.
\label{bofd}
\end{equation}
The right-hand side of Eq.~(\ref{gravprobd}) may look like a negative
probability density, but this is artificial since integration should be
done for decreasing $\delta$'s. We integrate Eq.~(\ref{gravprobd})
analytically and use that expression whenever numerical
values of the scattering probability to $\delta+\Delta\delta$ are
needed.

\begin{figure}
\center
\epsfig{width=\columnwidth,file=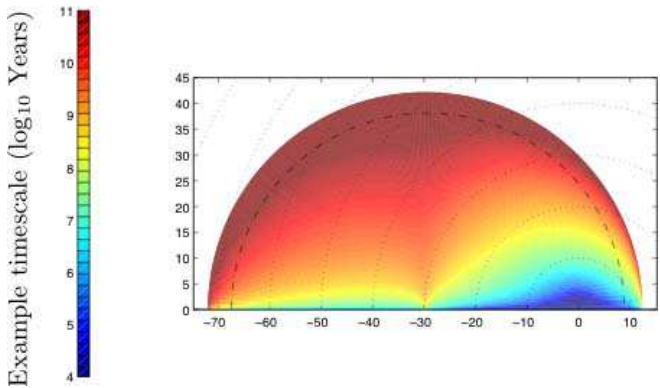}
\caption{{An example of the timescales of particle scattering.}
  The color bar indicates the time for which there is a $10$\%
  probability of scattering an angle $\delta=\pi/2\pm{}\pi/64$,
  depending on the present location of the particle. In general, time
  scales are shorter at lower velocities. By repeated close
  encounters with the Earth, particles may diffuse along the dotted
  circles (actually spheres) of constant velocity only, keeping $u$
  constant, but allowing for changes in $\theta$ and $\phi$. The
  figure shows the $\phi=75^\circ$ slice of the particle velocity
  space only.
  \label{pic_diff_tsc}}
\end{figure}

To get a feeling for the significance of the diffusion, we solve
Eq.~(\ref{gravprobd}) to obtain the typical time scales for scattering a given
angle to occur. As an example, we look at the 
time scales for which the probability of scattering with
$\delta=\pi/2\pm{}\pi/64$ is $10$\%. This is illustrated in Fig.~\ref{pic_diff_tsc}.

\subsection{The bound orbit density and orbit capture from the halo}

Let us now define \emph{the bound orbit density},
$n(\vect{u})$ to be the number of bound particle \emph{orbits} per
infinitesimal velocity and solid angle on the velocity sphere. The
orbit density is thus free from information about the particle location
along its elliptical orbit. The total number of bound particle orbits
in a thin shell of radius $u$ is
\begin{equation}
dN = du u^2 \int\!\!\!\int_{\Omega=\textrm{\small bound orbits}} 
d\Omega\: n(\vect{u}) \qquad 
\end{equation}
We will now divide each velocity sphere into cells (that can at this point be viewed as infinitesimally small).
The number of particles scattered between two locations on a sphere
of constant velocity in a given time must be an integral over the
source cell $i$ and the destination cell $j$,
$$ 
\frac{dN_{ji}}{dt} = 
\int\!\!\!\int_{ \alpha\in \Omega_i}\int\!\!\!\int_{\beta\in \Omega_j}
     \frac{dP(\beta,\alpha)}{dt} n(\alpha)
$$
In our case, the destination space is conveniently spanned by the
scattering angles $\delta$ and $\eta$.
The density of bound orbits scattered from cell $i$ to cell $j$ evolves with time as
\begin{equation}
\frac{d n_{ji}}{dt} =
\int\!\!\!\int_{\Omega_i} \! 
\ d\Omega 
\int\!\!\!\int_{K_j} \! d\delta \ \frac{d\eta}{2\pi} \  
\frac{dp(\vect{u},\delta)}{d\delta \: T_{\ert}} n(\vect{u}),
\label{boundtobound}
\end{equation}
where $K_j$ is defined to be the region in $\delta$--$\eta$--space
corresponding to scattering from the $i$ to the $j$ cell. The
scattering probability to the $[\delta,\delta+\Delta\delta]$ band is
evenly distributed over all cells in that region. Numerically this is
implemented as as loop over $\delta$ as measured from the center of
the source cell. The probability is then distributed over all discrete
cells whose centers are inside the current band.

It is important to understand that what we are considering is the
movement of the particle orbits, as opposed to the particles
themselves. This means that we do not need to calculate the actual
particle trajectories. When we are interested in the actual particle densities, we pick
orbits from the orbit densities. Note that there are two points on the orbit
that pass a given radius, but due to the
perihelion precession, any given particle could show up at any of these orbit
locations depending on the angle of perihelion. Since we have
anticipated mirror symmetry in the plane of the solar system, each
particle is smeared out at four indistinguishable orbit locations on
the sphere of constant $u$. This can always be done, regardless of the
existence of symmetries in the free distribution. If the free
distribution is not mirror symmetric in the ecliptic plane, it can
be forced to have this (in this case artificial) symmetry by averaging,
as long as we are only interested in the absolute capture of WIMPs in the Sun or planets.


The equations derived above apply only to particles which are already
gravitationally bound to the solar system. We now turn to the
calculation of the \emph{bound orbit density capture rate}; $\Delta
n_{j f}/T_{\ert}$ from the distribution of free particles. We will use
the local phase space density $F_f(\vect{u})$ [Particles$/(m^3\
m/s)$].

Consider the distribution of particles  $F_f(\vect{u})$ passing the Earth with
impact parameter $b$. The number of particles scattered an angle
$\delta(b\pm{}db/2)$ in a given period of time $T$, is
\begin{equation}
\underbrace{T u 2 \pi b db}_{dV} \: F_f(\vect{u}).
\end{equation}
According to Eq.~(\ref{deltaofb}) they are scattered an
angle $\delta(b)$. Using the relations (\ref{bofd}) we conclude that
the \emph{bound orbit density} at the cell $j$ will evolve with time as
\begin{widetext}
\begin{equation}
\frac{d n_{jf}}{dt} =
\int\!\!\!\int_{\Omega_\textrm{\small free}} \!\!\! \ d\Omega
\int\!\!\!\int_{K_j} \! d\delta \ \frac{d\eta}{2\pi} \  
\left( - \  
2 \pi u 
\frac{(M_{\ert} G)^2}{u^4}
\frac{\phantom{2}\cos\phantom{^3}(\delta/2)}{2
\sin^3(\delta/2)}
\: F_f(\vect{u}) \right),
\label{freetobound}
\end{equation}
\end{widetext}
caused by gravitational scattering from the halo. We now have
equations for gravitational diffusion as well as capture to the solar
system.

\subsection{Relating the 
phase space density $F(\vect{u})$ 
and the bound orbit density $n(\vect{u})$ }

The ideas of the last section can be used to write down an expression for
the phase space density, which is what we need for the weak capture
calculations. The relation between the phase space density
$F(\vect{u})$
and the \emph{bound orbit density}; $n(\vect{u})$
is derived as follows.

For a given orbit in the population of bound orbits, we use 
Eq.~(\ref{genyearprob}) to calculate the number of orbits that will
pass trough an area $\sigma$ each year. We now consider a volume $\textrm{d}V $ in
space with base area $\sigma$ and height $h$ such that $h$ is parallel
to $\vect{u}$. A particle passing trough the area will spend a time $h/u$
in the volume. This means that the fraction of the WIMP year spent in
the volume in case of an encounter is
$$
 \frac{h}{u T_\chi}.
$$
The fraction of orbits passing through $\sigma$ each  WIMP year is
$$
\frac{p(\sigma,\vect{u})}{T_{\ert}} T_\chi
$$
Therefore, since $F(\vect{u})$ is the number of particles per $du^3
dV$, the relation between $F(\vect{u})$ and $n(\vect{u})$ is
\begin{equation}
 F(\vect{u}) \textrm{d}V   = n(\vect{u}) 
{\frac{h}{u T_\chi}} 
{\frac{p(\sigma,\vect{u})}{T_{\ert}}
T_\chi.}
\end{equation}
or
\begin{equation}
F(\vect{u}) \textrm{d}V  = n(\vect{u}) \: 
h \: \frac{ p(\sigma,\vect{u})}{ u T_{\ert}} = %
\: n(\vect{u}) \frac{3}{2}\frac{ \textrm{d}V  }{\pi R^2}\frac{v_{\ert}}{u^2} \gamma(\vect{u})^{-1}
\label{freln}
\end{equation}
One should note that by construction $F(\vect{u})$ above is valid in the
frame of the planet. However, the right hand side of the
equations above presumes the planet to have a constant velocity during
the encounter so that $\textrm{d}u^3$ are equal to the velocity volume
element in the frame of the Sun, $\textrm{d}w^3$.

\section{Solar depletion of bound orbits}
\label{sec:soldepletion}

In the previous section, we investigated the evolution of the bound orbit densities due to
scatterings from other bound orbits and from free orbits. We have one main piece remaining to be studied, and that is the effects of solar depletion, i.e.\ how much of the bound WIMPs are actually captured by the Sun, thus reducing their density in the solar system.
 
We have done this by numerically calculating the actual motion for different WIMP orbits in the solar system over 49 million years. As a measure of the quality of the numerical methods, we have also calculated the fates of the 47 asteroids studied by Farinella
et al.\ \cite{farinella}, as presented in appendix \ref{app:farinella}.

\subsection{The numerical methods and conditions}

We have numerically integrated the orbits of about 2000 particles in
typical Earth crossing orbits in order to estimate the solar depletion.
The particles were spread out on the bound velocity space with
random initial positions on the Earth's orbit. 
We have mainly used the \textsf{Mercury} package \cite{merc} by Chambers for the integration.
It has the most important numerical algorithms, such as Everhart's
15:th order \textsf{Radao} \cite{radao} with Gauss--Radao spacings,
and the equally well-known \textsf{Bulirsch--Stoer} \cite{bulstoer} algorithm.
Both are variable step size algorithms dedicated to many body problems,
and are commonly used in asteroid research for problems similar to
ours. The package also includes a set of symplectic algorithms, which
have been used for some tests.
By looking at some test orbits, we found that the symplectic
algorithms (at least as implemented in the \textsf{Mercury} package) were slower and less accurate for our setup. The tested
symplectic algorithms were ''MVS: mixed-variable symplectic'' \cite{wisdom}
as well as ''Hybrid symplectic/Bulirsch-Stoer'' \cite{merc}.

The calculations included the test particles, the Sun, the Earth,
Jupiter and Venus. Other planets were not included as they are believed to be sub-dominant. The \textsf{Bulirsch--Stoer} algorithm was used to calculate
the orbits of all test particles, as well as the planets, during a time of
49 million years. This took about 35\,000 CPU hours, on a variety of Linux and 
Alpha machines.
A wide range of different accuracy parameters were used, from
$10^{-14}$ to $10^{-8}$, to evaluate the role this plays. The
numerical representation of the real numbers limits the benefit of
going past about $10^{-12}$. The final choice of $10^{-10}$ is a
balance between time and accuracy.
A recent publication \cite{migrjup} in the subject of
numerical simulations of a special set of Jupiter crossing asteroids,
came to a similar conclusion; When using the Bulirsch--Stoer algorithm
for their calculations, they found accuracy parameters in the range of
$10^{-9}$--$10^{-8}$ to give statistically similar results as
$10^{-12}$.
In the comparisons carried out, this gave results very similar to those with
higher accuracy parameters. The comparison with the
\textsf{Radao} algorithm gave qualitatively similar, however not
identical, results with a similar calculation speed. In some occasions
however, the \textsf{Radao} algorithm gave a higher solar depletion for
particles with very high velocity (relative to the Earth), $u \gtrsim 50$ km/s. This is not
of much concern for our purposes though as we are mainly interested in much lower velocities for Earth capture to be efficient.

For ordinary asteroid calculations, a point mass approximation combined with collision
detection is sufficient. Our case is a little more delicate since
WIMPs may pass through the planets. To handle this,   
the gravitational routines were modified to use the real
gravitational potentials inside the planets.

For Jupiter and Earth, we used ''true'' mass distributions
\cite{Jupitermass,EncBrit}. For Venus we rescaled the mass
distribution of the Earth and removed the liquid iron core. Other
planets included in tests where assumed to be homogeneous. The
improvement allows the particles to pass through the planets without
being infinitely scattered by a point mass, making the calculations
more realistic and numerically stable. For completeness, it would be
interesting to add more planets to the simulations, but it is unfeasible
to do as it slows down the calculations too much. We also believe, that we have included the most important planets in our simulations.

\subsection{The results of the numerical simulations}

\begin{figure}
\center
\epsfig{width=\columnwidth,file=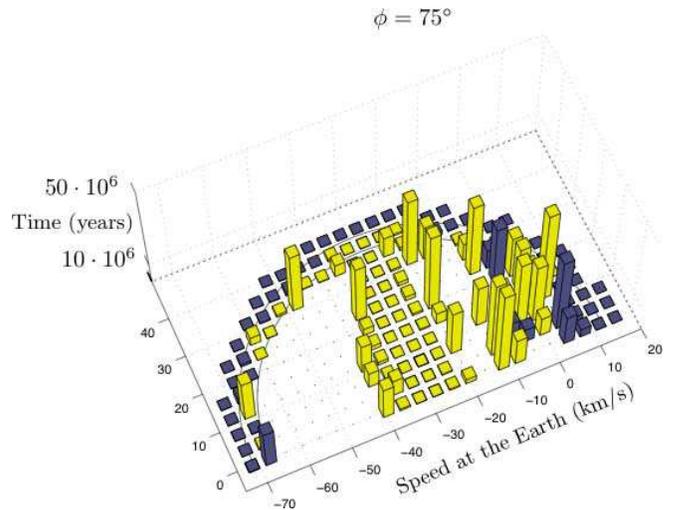}
\caption{{The time (linear scale) for ejection (blue/dark gray)
    and capture in the Sun (yellow/light gray) of a set of test
    particles.} Each bin represents only one particle, so the
    statistical error is high. However, this figure is typical for all
    angles, except that the plateau of fast solar depletion at large
    ''backward'' velocities are raised when $\phi$ approaches
    90$^\circ$. Some particles survived in the Solar system for the
    whole of the simulation. Those particles are marked with black dots.
 \label{fig_soldep15}}
\end{figure}

\begin{figure}
\center
\epsfig{width=\columnwidth,file=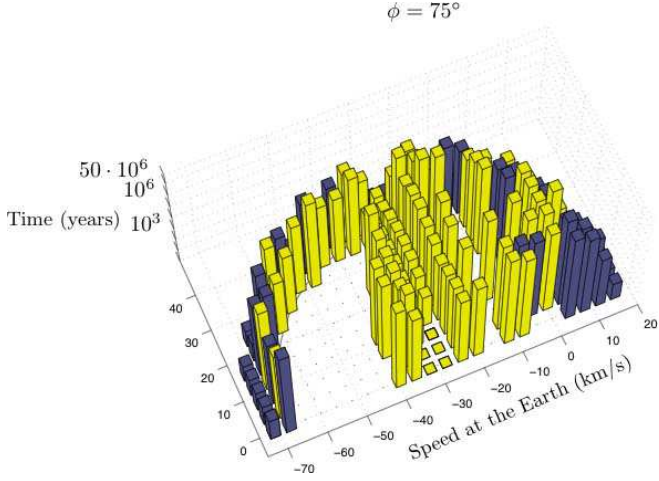}
\caption{{The time (log scale) for ejection and capture in the Sun of
    a set of test particles.} This figure is identical to figure
    \ref{fig_soldep15}, except for the time scale which is
    logarithmic. In this scale, it's easier to see that there is a
    small region at $-30$ km/s where the solar depletion occurs
    directly. This is not surprising, since this region corresponds to
    particles with very low velocity \emph{in the frame of the Sun}.
    The plateau of direct solar capture extends further in the special
    case of $\phi=90$ (not shown) which allows
    extremely elliptic, or radial orbits. (The plane of start
    positions is then parallel to the ecliptic plane.)
 \label{fig_logsoldep15}}
\end{figure}

The solar depletion was mainly calculated for particles in eight
planes of $\vect{u}$ space, with the $\phi$ values $0, 15, 30, 45, 60,
75, 90$ and $-30$ degrees (the $\phi=-30^\circ$ plane was used to
investigate the expected radial mirror symmetry of the results). Our
solar depletion results are not as bad as Gould feared \cite{gould-conserv}; 
Most of the particles survived two million years. Nevertheless, solar capture is too
large to be ignored. Figs \ref{fig_soldep15} and
\ref{fig_logsoldep15} show the $\phi=75^\circ$ plane, and the times
after which the particles hit the Sun. We note that ejection is much more
common at Jupiter-crossing orbits. This is in compliance with the fact
that, according to the scattering model used here, the
probability of scattering for such orbits is high. 
The fact that there is a large region at $-50$ km/s where there are no
ejections or sun captures, is in agreement with the qualitative
results by Gould, presented in his 1991 paper \cite{gould-diff} (see
his Fig.~3, where he assumes that the filling times are about the
same as the time of ejection). Apart from the calculations shown here,
some extra calculations were carried out for relative velocities lower
than $15$ km/s. The results of those calculations were incorporated
and used in the same way as the others.

Another important, however simple, result is that there seem to exist
a mirror symmetry in the in-out directions. This is expected, since
particles may hit the Earth both on its way out and on the way back on
its perihelion revolution. Considering Fig.~\ref{pic_preiorb}, it's
evident that this is equivalent to a symmetry in the sign of $\phi$
claimed by Gould; that the $\phi$ and $-\phi$ cases are identical.

\begin{figure}
\center
\epsfig{width=\columnwidth,file=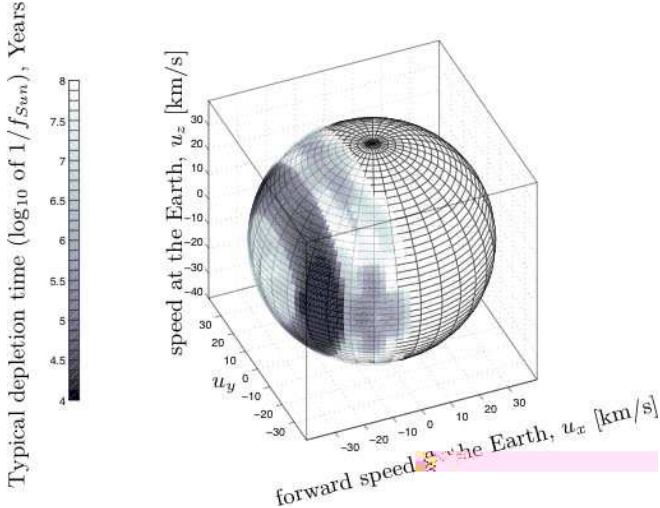}\\[2ex]
\caption{{The solar depletion at the $u=40$ km/s sphere.} The
color bar indicates the logarithm of the typical depletion time
$1/f_{Sun}$. The region to the right are the free orbits, for which
the solar depletion is irrelevant. At a ''backward'' velocity of
$30$km/s, the Sun-depletions is greater, in agreement with the
previous figures of this section. In understanding this figure, it may
help to take a look at the $u=40$ km/s line of figure
\ref{pic_diffprincip}, which corresponds to
the central horizontal ($\phi=90^\circ$) plane of this figure.
\label{fig_deplonsphere}}
\end{figure}

The particle orbits were evenly distributed in velocity space, but we solve the diffusion equations on spheres of constant $u$, hence we interpolate our results.
What we need to extract from our numerical simulations is the depletion frequency, i.e.\ the expected depletion probability per given time.
Since the form of the actual distribution, of which the results
of the numerical calculations are samples, are unknown, the most
reasonable way to estimate the depletion probability per unit time is
\begin{equation}
f_{Sun}=\frac{1}{T_{Sun}}.
\end{equation}
Figure \ref{fig_deplonsphere} shows the logarithm of $1/f$, interpolated onto
a sphere of constant $u$, namely $u=40$ km/s.

\section{The evolution equations for one planet}
\label{sec:nummethods}

In the previous sections we have presented the analytic expressions for the scattering of bound orbits to other bound orbits, Eq.~(\ref{boundtobound}), as well as capture from free to bound orbits, Eq.~(\ref{freetobound}). We have also, by numerical simulations, estimated the rate at which orbits are sent into the Sun and thus captured. 
We are primarily interested in how the bound orbit density evolves with time, and will here write down the dynamic equations in a form suitable for numerical work.

\subsection{The dynamic equations of the bound orbit density}

The bound orbit density develops in time in the following way,
\begin{eqnarray}
  \lefteqn{du\;u^2 
    \frac{d n_{j}}{d t} =   
    du\;u^2
    \Bigg[ 
    \sum_{i\in\textrm{bound}} \left( \frac{d  n_{ji}}{d t} -
    \frac{d n_{ij}}{d t} \right)} 
    \nonumber \\
    & & + \sum_{f\in\textrm{unbound}} \left(\frac{d n_{jf}}{d t} - 
\frac{d n_{fj}}{d t}
    \right ) -
    \frac{d n_{sj}}{d t} 
    \Bigg]\!,%
\label{orbdevel}%
\end{eqnarray}
where $n_j$ is the number of orbits in the small cell
\footnote{Note that if $du$ is not small, the volume of the cell is not $d
\Omega u^2 du$, but $\Omega u^2 du(1+{du^2}/(12 u^2))$. When numerical values are called for,
we substitute $du$ with $\Delta u(1+{\Delta
u^2}/(12 u^2))$.}
 $j$ of the
sphere. The sum over $i$ is the flow from and to the other bound
cells. The $n_{jf}$ and $n_{sj}$ terms are representing capture from
unbound orbits and capture of bound orbits by the Sun, while the
$n_{fj}$ term represents the ejection of bound particles.

We will now reformulate Eq.~(\ref{orbdevel}) in matrix form suitable for numerical calculations. Let us first define our state vectors,
\begin{equation}
X = \left(\!
      \begin{array}{c}
        N_{s}               \\
        n_{i}               \\
        F_{f}               \\
      \end{array}\!
\right)
\end{equation}
where $N_{s}$ is the number of particles captured by the Sun, $n_i$ is the bound orbit density and $F_{f}$ is the velocity number density of free (unbound) orbits.  
If the cells $i$ are small enough, the various densities can be considered constant over each cell. Using this and the fact that the
$\eta$ part of the integration 
is independent of $n(\vect{u})$ and
$F_f(\vect{u})$, this means that Eqs.~(\ref{boundtobound}) and
(\ref{freetobound}) can be written as
\begin{eqnarray}
\frac{d n_{ji}}{dt } & = & p^{\rm bb}_{ji} n_i(\vect{u}) \quad\textrm{and}
\label{miniboundtobound} \\
\frac{d n_{jf}}{dt } & = & p^{\rm bf}_{jf} F_f(\vect{u})\phantom{  \quad\textrm{and}} \label{eq:dn2}
\end{eqnarray}
The solar capture can be written in the same way:
\begin{equation}
\frac{d n_{si}}{dt } = p^{\rm sc}_{si} n_i(\vect{u})
\label{matrisekv_sun}
\end{equation}
The $p$:s can be considered as the probability per unit time to transfer particles/orbits from and to the various cells. A positive $p$ means that we transfer to the cell and a negative $p$ that we transfer from the cell. The $p^{\rm bb}_{ii}$ element requires an explanation. This is the probability per unit time that an orbit in cell $i$ is not scattered to another bound or free cell, i.e.\ this term includes all the scattering out to both other bound orbits, and unbound orbits. The probability for solar capture though is handled separately by $p^{\rm sc}_{si}$. As the various entries in the state vector $X$ have different units ($N_s$ is a number, $n_i$ is the the orbit density and $F_f$ is the number density), the required conversion factors are also included in the $p$:s. The cells can be of various size, and these sizes are also included as weights in the $p$:s. We will not write down explicitly the expressions for the $p$:s as they are found elsewhere: $p^{bb}$ and $p^{bf}$ can be extracted from \ Eqs.~(\ref{boundtobound}) and (\ref{freetobound}) respectively, while $p^{sc}$, on the other hand, we extract from our numerical simulations of solar capture, discussed in section \ref{sec:soldepletion}. E.g.\ the $p^{sc}$:s for the cells on the 40 km/s sphere can be read off from Fig.~\ref{fig_deplonsphere}.

Integrating Eqs.~(\ref{miniboundtobound})--(\ref{matrisekv_sun}) over time, and replacing $dt$ with a discrete time step $\Delta t$, we can write the evolution of the state vector $X$ as
\begin{eqnarray}
  \lefteqn{X(t_0 + \Delta t) 
   =  \vect{T}(\Delta t) X(t_0) \quad \mbox{, with}} \nonumber \\
& & 
\vect{T}(\Delta t)  =  \left(\!
      \begin{array}{c c c }
      1  &  \vect{P}^{\rm sc}               & \mathbf{0}             \\
      \mathbf{0}  &  \mathds{1} + \vect{P}^{\rm bb} (\mathds{1} - \vect{P}^{\rm sc}) - 
      \vect{P}^{\rm sc}   & \vect{P}^{\rm bf}  \\
      \mathbf{0}  &  \mathbf{0}           &   \mathds{1}     \\
      \end{array}\!
\right).
\label{matris_ekv}
\end{eqnarray}
The first row describes the solar capture of bound orbits. The
second row describes the development of bound orbits, and the capture
of free orbits. Its height is given by the number of bound cells. The
last row is a little bit special. One may propose that bound WIMPs
scattered to unbound orbits should give a contribution in the second
column. However, such particles will not meet the Earth again, so the
lowest part of the matrix should only do the job of keeping the
unbound phase space density constant. The size of the last row unit
matrix is of course given by the number of free orbit cells $F_f$. The matrices $\vect{P}$ can  be regarded as transition probabilities (for the given time interval $\Delta t$).  The elements in the $\vect{P}$ matrices are given by
\begin{eqnarray}
P^{\rm bf}_{if} (\Delta t) & = & p^{\rm bf}_{if} \Delta t \\
P^{\rm bb}_{ij} (\Delta t) & = & p^{\rm bb}_{ij} \Delta t \\
P^{\rm sc}_{ij} (\Delta t) & = & \delta_{ij} p^{\rm sc}_{si} \Delta t 
\end{eqnarray}
for free to bound orbits, bound to bound orbits and solar capture respectively.

\subsection{The bound orbit density at arbitrary times}

Eq.~(\ref{matris_ekv}) describes the evolution of the state vector $X$ during a time step $\Delta t$. 
We can write the time development operator that takes us to any time $t$ as 
\begin{equation}
  \vect{U}(t) \equiv \left[ \vect{T}(\Delta t)\right] ^{t/\Delta t} \qquad X(t_0+t)=\vect{U}(t) X(t_0)
\end{equation}
The exponentiation of $\vect{T}$ can be done either by diagonalizing $\vect{T}$, or (for applicable times $t$) by repeatingly quadrating $\vect{T}$.
We have calculated and diagonalized $\vect{T}$'s with a variety of different
cell configurations. A simple polar grid is a good first choice, but
it has a large spread in shape and area of the cells, which means that
valuable memory and calculation time is wasted. Therefore, we have
used cells with the shape of spherical triangles, built from
icosahedrons or octahedrons. The cells of the body were successively
divided in four nearly identical spherical triangles, until the right
number of cells were reached. The velocity space of each planet was
built up of about 65 spheres, usually with 2048 cells each, which means a
total of about 130\,000 discrete cells for each planet.

\begin{figure}
\center

\includegraphics[width=0.30\columnwidth]{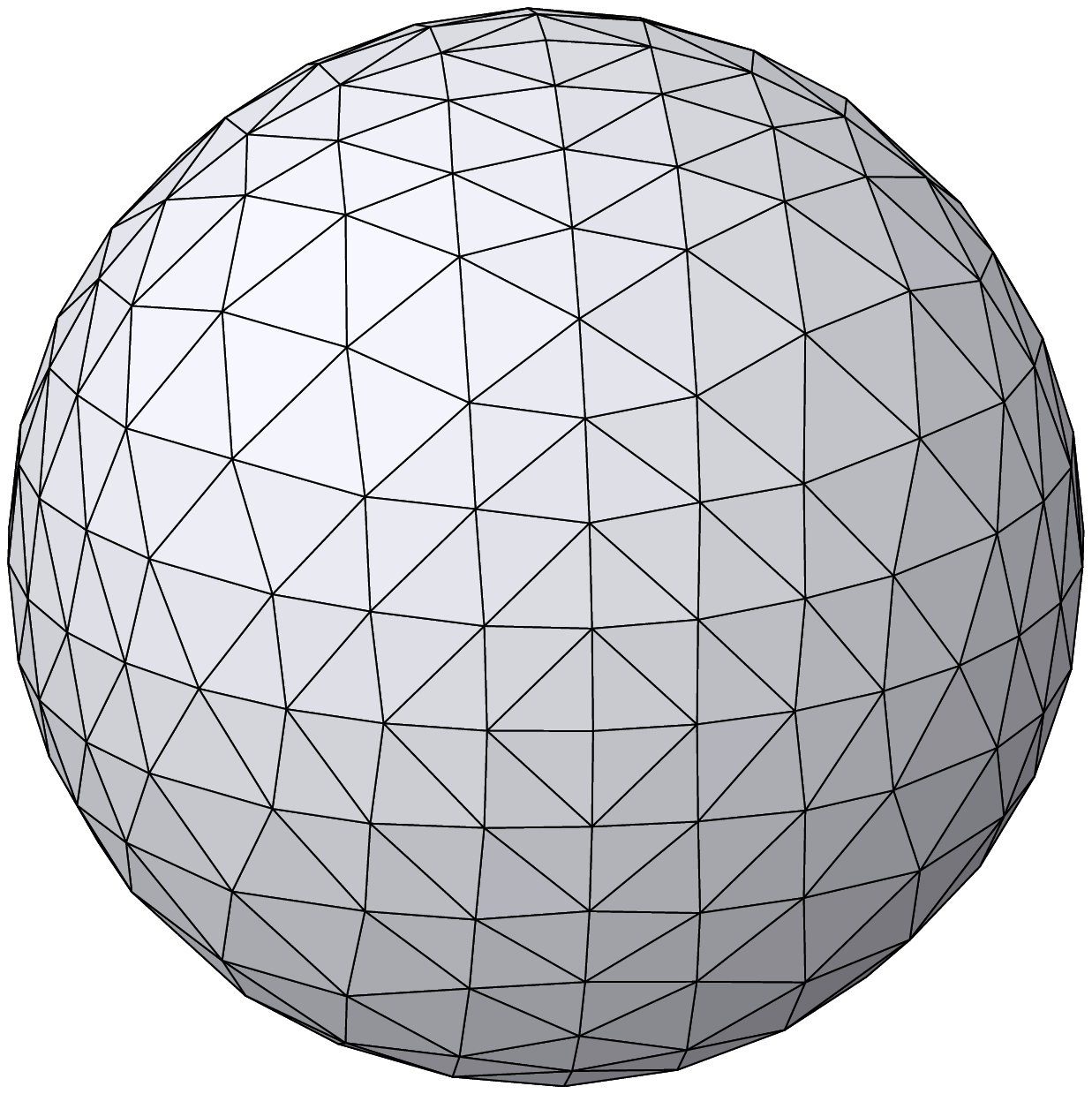}
\includegraphics[width=0.30\columnwidth]{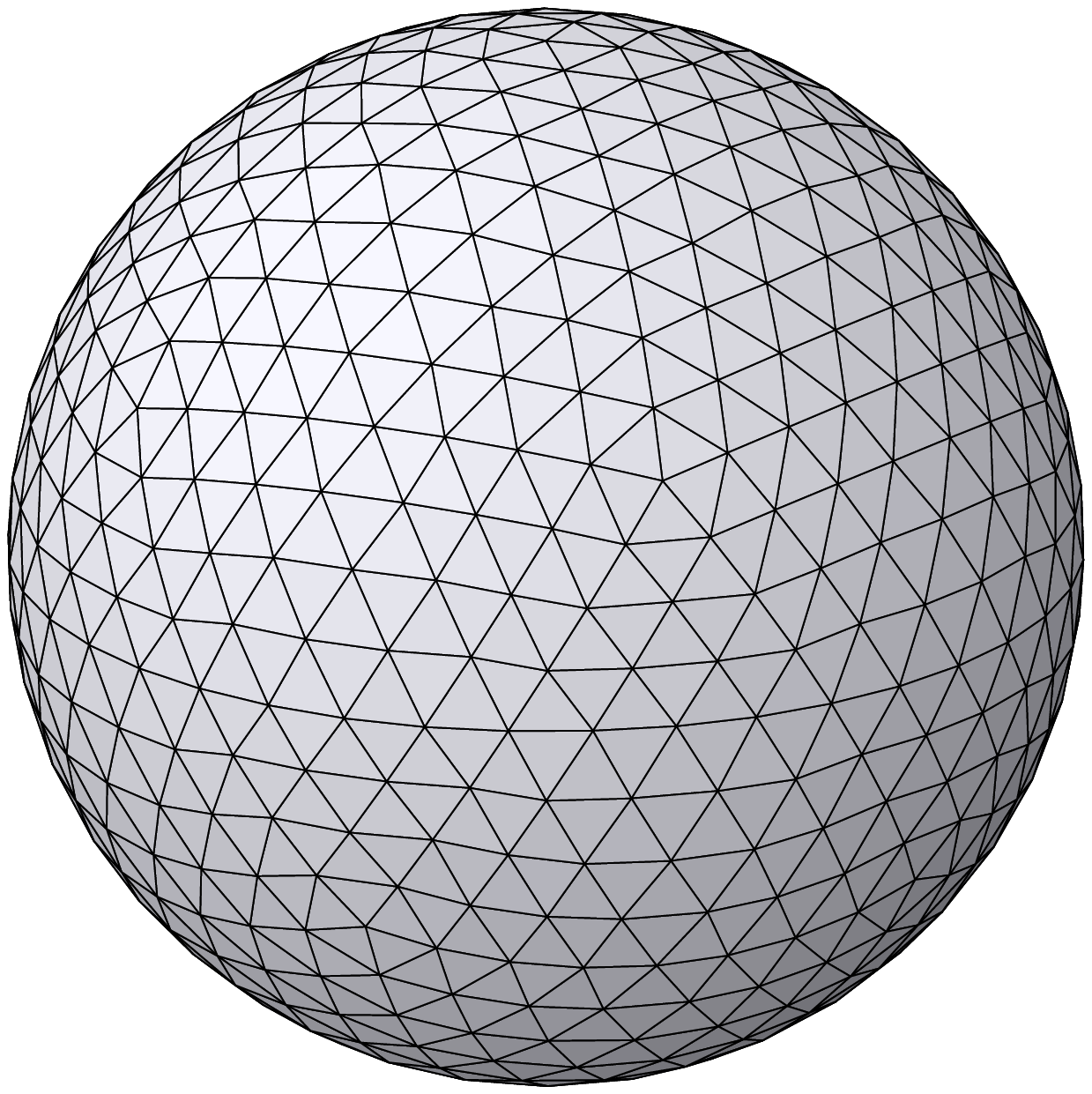}
\includegraphics[width=0.30\columnwidth]{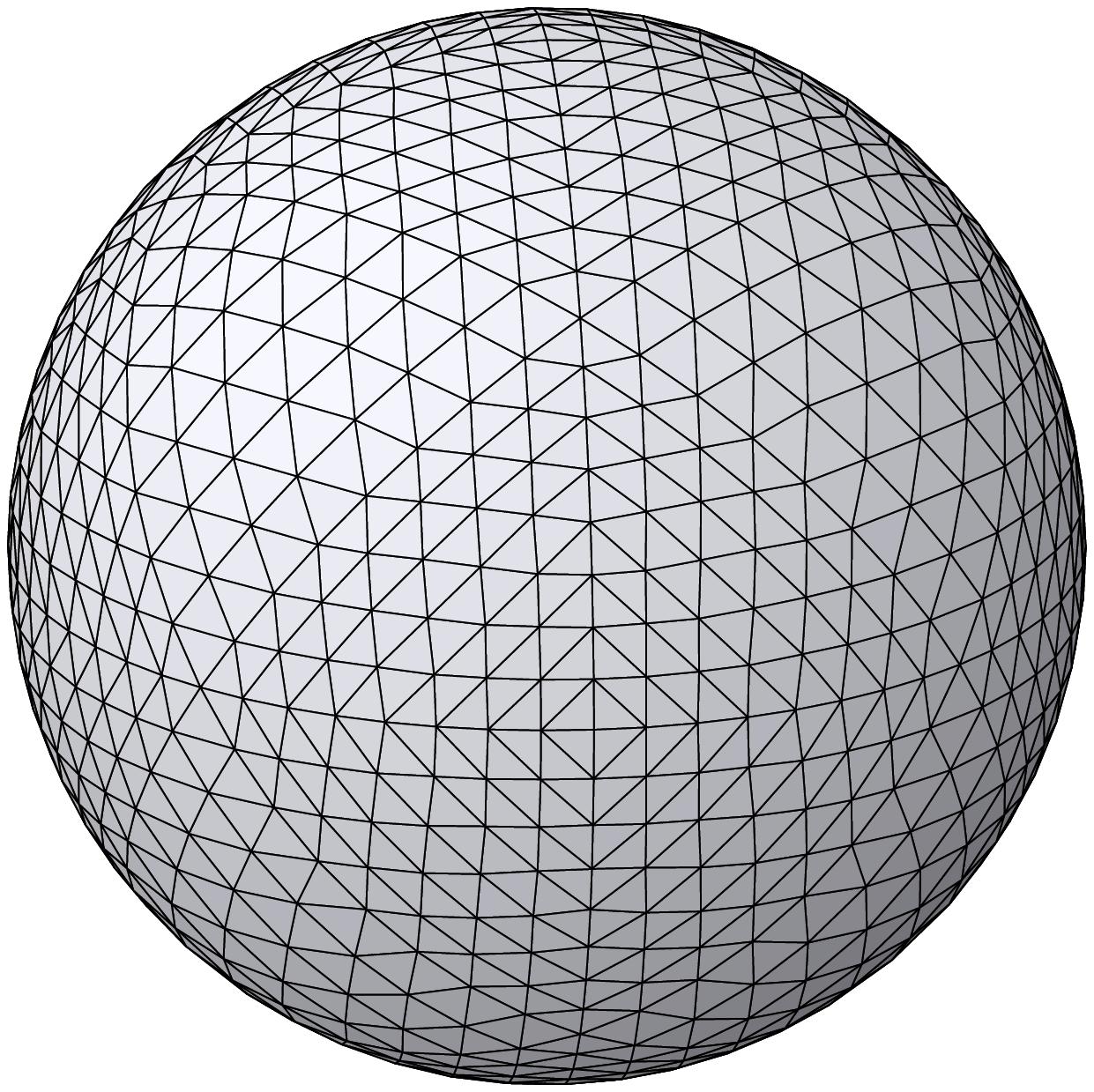}
\caption{The discretization of space were made using triangles. The
number of cells on the displayed triangles are 512, 1280 and 2048, of
which the last one was used in our final calculations.
  \label{pic_trianglecells}}
\end{figure}

If the octahedron is used as a starting object, it's possible to rotate
the sphere to obtain mirror symmetry in the in-out (radial in the solar
system) and up--down directions. Since the problems to solve possess
the same symmetries, this reduces the size of the state vectors by a
factor of four, and the time evolution operators by a factor of 16.

Most of the numerical calculations take place in this
compressed space. By making this a run-time option, we have verified
that this does not introduce any errors. Great efforts
have been put in verifying the consistency of the time evolution
operator. As a simple example, the probability for a particle to end up
\emph{anywhere} is unity. Making use of the mirror
symmetry of the equations, it has been possible to calculate and
diagonalize $P$'s with up to some $10^8$ elements. The set of
time evolution operators of a planet, can be thought of as one huge
block diagonal time evolution operator for the 130\,000-dimensional cell
space. The block shape of the matrix is dependent of the conservation
of energy when a particle is repeatingly scattered by a single planet.
We have also checked the robustness of our results
as the number of cells varies. If one uses too few cells, one would expect that the effect of diffusion is underestimated as small-angle deflections (smaller than the cell size) are then artificially suppressed. At velocities above 8 km/s, our results do not change significantly when going from 1024 to 2048 cells. Below 8 km/s, however, the resulting WIMP density is somewhat larger in our simulation with 2048 cells than in our simulation with 1024 cells. This would indicate that the density at these low velocities could go up somewhat if we used even more cells. However, it is not possible to increase the number of cells further, as it is only feasible to perform these simulations when the full velocity space can be maintained in the computer memory simultaneously. It is also not reasonable to perform this part of the calculation more accurately than other parts, like the solar capture discussed in the previous section.

We have now set up a framework for diffusion from one planet. We have done this following
the scheme set up by Gould \cite{gould-diff}, with some small modifications and improvements. Our main goal has been to make it possible to include the effects of solar depletion, and hence we have formulated the diffusion problem in a form suitable for numerical work, where the inclusion of solar capture is easily done.

In the next section we will put all of this together, where we also include the diffusion effects of the other (dominant) planets.

\section{The velocity Distribution at the Earth: Combining the
  effects of Jupiter, Venus and the Earth}

\label{sec:combined-diff}

We have so far considered the diffusion caused by one planet at a time and the effect of solar capture. We are now ready to include more than one planet in our treatment. In section \ref{sec:one-diff}, where we investigated the diffusion effects caused by one planet, we saw that one planet can only change the direction and not the velocity of a WIMP. However, WIMPs that have different directions, but the same velocity at one planet, will not only have different directions, but also different velocities at another planet. Hence, the main effect of including more planets in the diffusion is to diffuse particles also to different velocities. We thus have a mechanism to populate a larger part of the phase space at Earth, and this process is hence very important, especially for heavier WIMPs.
We will here include the diffusion effects of Venus, the Earth and Jupiter as these are the planets dominating the diffusion mechanism \cite{gould-diff}.

\subsection{Transformation of coordinates and bound orbit density when changing planet}
\label{forstasect}

The velocity and angles in a planet-based coordinate system at a
planet with orbit radius $a$ and velocity $v$, can be converted to the
coordinates of another planet via the energy, angular momentum and
inclination. This is not enough for the specification of the exact
location of the particles, but we are only interested in the shape and
orientation of the \emph{orbits},
\begin{eqnarray}
  E &=& \frac{1}{2}(u^2 + 2 u v \cos\theta + v^2 - 2\frac{M_\odot G}{a})\\
  L &=& a(v+u \cos\theta ) \\
  \tan i &= &\frac{u \sin \theta \cos \phi}{v+u\cos\theta}
\end{eqnarray}
The inverse transformation is
\begin{eqnarray}
 u^2   &=& 2 ( E -  L \frac {v}{a} + \frac{1}{2} v^2 + \frac{M_\odot G}{a}) \\
\cos \theta &=&  \frac{1}{u}   \left(\frac{L}{a} -v \right), \\
\cos \phi   &=&   
                \frac{ L \tan i }{ a u \sqrt{ 1 - \cos^2\theta }}
\end{eqnarray}
As an example, we will transform the various densities, as seen in
the frame of the Earth to the corresponding quantities at Venus.

The change of frame consists of two Galileo transformations, as well
as the journey of the particles in the potential force of the Sun.
Since the first is just a change of origin in the 6-dimensional phase
space, the change of frame obeys Liouville's theorem,
\begin{equation}
 F_{\ven}(\vect{u_{\ven}}) = 
 F_{\ert}(\vect{u_{\ert}})  
\end{equation}
Using Eq.~(\ref{freln}), the orbit densities at the two locations
can now be related as 
\begin{equation}
 n_{\ven}(\vect{u_{\ven}}) = 
 n_{\ert}(\vect{u_{\ert}}(\vect{u_{\ven}}))  \ J_{{\ert}
 \ven}(\vect{u_{\ven}})\textrm{, where}
\label{planchange}
\end{equation}
\begin{equation}
 J_{{\ert} \ven}(\vect{u_{\ven}}) = \frac{v_{\ert}}{v_{\ven}}
\left( \frac{R_{\ven} u_{\ven}\phantom{(\vect{u_{\ven}})}}{R_{\ert}
u_{\ert}(\vect{u_{\ven}})}\right)^2
\frac{\gamma_{\ven}(u_{\ven})}{\gamma_{\ert}(u_{\ert}(\vect{u_{\ven}}))}
\label{eq:jacobian}
\end{equation}
is the Jacobian.

\begin{figure}
\center
\epsfig{width=\columnwidth,file=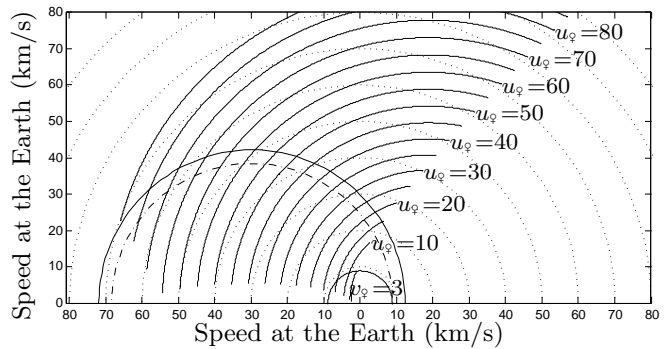}
\caption{{A detailed view of the $\phi=75^\circ$ slice of the particle velocity space
  at the frame of the Earth.} The solid archs represent particles in
  Venus--crossing orbits, and their speed \emph{at the location and in
  the frame of Venus}. Since the archs are part of spheres of constant
  velocity at Venus, they show the possible directions of diffusion
  caused by Venus. Jupiter crossing orbits (not shown) fill the region
  between the dash--dotted and the solid semicircles. The low velocity
  region of this figure will be discussed around figure
  \ref{pic_ven_jup_small} in section \ref{venus_jorden_princ}.
  \label{pic_ven_jup}}
\end{figure}

Using these transformations, it is possible to
investigate how a sphere of constant velocity at a specific planet
will look when the particles pass the Earth. Figure \ref{pic_ven_jup}
is an example of this. 

The archs of constant velocity \emph{in the
frame of Venus} are shown to indicate the directions of diffusion
caused by that planet. Since the lines of constant velocity at Venus
cross the $u_{\ert}$=12.3~km/s line, Venus may diffuse particles into
the important $u_{\ert}<$12.3~km/s region.

\subsection{Solving the many body diffusion problem}
 
\label{sec:solvemany}

Consider a point in velocity space, in the frame of the Earth,
\emph{and in Jupiter-crossing orbit}. At this point, the bound orbit
density $n(\vect{u})$ takes on a value $n_A$ at a given time $t_0$.
Call this density, transformed into a specific point in the reference
frame of Jupiter $n_B$. Now, after a short period of time, from now on
called \emph{step size}, the Earth may have increased (or decreased)
$n_A$ by an amount $d\,n_A$, and Jupiter may have increased (or
decreased) $n_B$ by $d\,n_B$. Since $n_A$ and $n_B$ are really a
measurement of the same density, and there are two processes
(interactions with the two planets) affecting the the differences, 
the orbit densities after the time step are given by
\begin{eqnarray}
\begin{array}{r l}
n_A \rightarrow  & n_A' = n_A +  d\,n_A  +  J  d\,n_B
\\[0.5em]
n_B \rightarrow  & n_B' = n_B +  d\,n_B  +  J^{-1} d\,n_A
\end{array}
\label{interpABBA}
\end{eqnarray}
where $J$ is the Jacobian for the transformation. Note that the step size introduced above is the step size after which transfer of densities between planets occur. For the diffusion effects of the individual planets during this step size, we use much smaller time steps.

In order to transfer the orbit densities from one planet to another in
a numerically reasonable way, all cells at each planet is matched
to the correct cells on the other planet. Since there is not a one to one correspondence between the cells of different planets, we need to interpolate between cells. 
We use a linear interpolation, but have also checked that a simpler nearest neighbor 
interpolation gives similar (but more noisy) results.

The velocity spaces of all pairs of involved planets were tessellated,
in order to create the matrix of linear interpolation. This means that
each cell was identified to constitute the corners of to up to six
octahedrons. All transformed points were identified to belong to a
single octahedron, and the location of the transformed point was given
as a linear combination of the octahedron corners. This linear
combination was then used as interpolation for the densities.

\subsection{Numerical issues}

In the previous subsection, our scheme for taking care of the diffusion effects of more than one planet was outlined. We will here discuss the measures we have taken to make sure that
our numerical implementation is stable and does not introduce numerical artifacts.

In order to further improve the stability of the interpolation between the planet cells, the
orbit densities $n$ are never interpolated directly. Instead, all
interpolations are done between phase space densities $F$, and
then converted to the $n$--space of the respective planets. The phase space
density $F$ is a slowly changing function, while
$n$ is not. This is so since among other things, the roughness of
$\gamma$, Eq.~(\ref{gammaekv}), is inherited to $n$, but removed
again when $F$ is calculated.

At any time, the densities at the two planets must be
consistent with each other so that a density at a particular point in
one frame matches that of the point transformed to the other planet,
as described by Eq.~(\ref{planchange}). Small interpolation errors can build up with time though, and we need to take care of this potential problem. To force the densities at the two planets, $n_A$ and $n_B$ to be
consistent, they were regularly averaged as follows:
\begin{equation}
\begin{array}{r l}
( n_A' + J n_B')/2 \rightarrow & n_A'' \\[0.5em]
( n_B' + J^{-1} n_A')/2 \rightarrow & n_B''
\end{array}
\label{interpABBAavrg}
\end{equation}
From an analytical point of view, this is not needed, but it turns out
to be a good way of making the algorithm more numerically robust.

The results are stable with respect to step size as well as shape of
the velocity space used. This is particularly true when the averaging outlined above is
done. It is not necessary to perform the averaging after every time step, instead we can perform it much more seldom. Even if we perform the averaging for all velocity spheres, it turns out that it is unimportant in the region above $u\approx$10 km/s, where
the processes are slower and stable anyway.  In the steep region below $u=7$ km/s, averaging is needed though to keep the stability. We have verified that in the limit of very small step sizes, the unaveraged results approach the ones with averaging even in this region, but averaging allows us to get better accuracy and stability even with longer step sizes.
We have also verified that the results are quite
stable with respect to the averaging frequency. The result figures of the previous
section show the results of a small step size; 16 thousand years.

A related problem is that even though the
Jacobian determinant of Eq.~(\ref{eq:jacobian}) is
mathematically valid, linear interpolations do not assure conservation
of mass. This means that when repeatingly transferring density
information between a pair of planets, one can not be sure that the
interpolation does not, in error, introduce or remove mass from the
system. These artificial 'sources' or 'sinks' need to be removed.
While it is not possible to do this on a cell by cell basis, we have
investigated the total mass transferred to and from each planet and used this to renormalize the mass transferred to ensure mass conservation. In
equilibrium, the error is quite small; under one percent, but when a distribution is built up, the error can be larger than that.

We have further tested that the step size is not critical for the results. This indicates that
our numerical implementation, with the stability measures outlined above, is stable and that the possible errors are under control.

We have now presented two methods to keep the possible numerical artifacts under control:
the averaging process to keep the densities consistent between the planets and the renormalization process to force mass conservation. Both of these make our algorithms both more stable and reliable.

\subsection{Investigation of Jupiter--crossing orbits}
\label{earthjupvensys}

It turns out, that the density of Jupiter-crossing orbits is
independent of the diffusion effects of the Earth as well as those
of Venus. This is expected, since the mass of Jupiter is so much
larger, and the scattering probability increases with the planet mass
squared, see Eq.~(\ref{gravprobd}).

To investigate this, we have numerically solved the Earth--Jupiter
diffusion system in two ways: calculating the evolution with Jupiter alone, as
well as solving the two body diffusion problem with the methods
described above.
In either case, it takes only a couple of million years for Jupiter's
Earth crossing orbits to come into equilibrium with the unbound orbits.
This means that for Jupiter--crossing orbits we can safely neglect the diffusion effects of the other planets and let Jupiter fill these orbits alone.
It also turns out that the diffusion of Jupiter--crossing orbits is so much faster than solar depletion, and we can thus ignore solar depletion for these kind of orbits.

We can then already now see that the ultra conservative view in Gould and Alam \cite{gould-conserv} is too pessimistic and that at least as many bound WIMPs as in the conservative view remains in the solar system. We will next see what the fate is for bound orbits further inside the solar system.

\subsection{Investigation of the Earth--Venus--Jupiter system}
\label{sistasect}
\begin{figure}
\label{venus_jorden_princ}
\center
\epsfig{width=\columnwidth,file=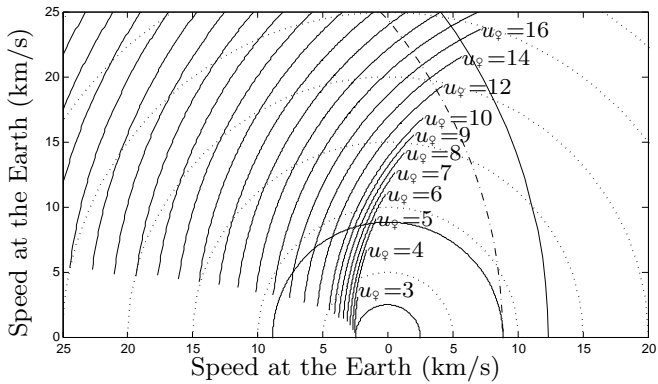}
\vspace*{-2em}
\caption{The $\phi=75^\circ$ slice of the particle velocity space
  at the frame of the Earth, the low-velocity region. The region
  outside the solid $u\approx9$ km/s line is populated by the combined
  effect of Jupiter and the Earth. The diffusion effect of Venus is needed for particles to
  reach the region between the $9$ km/s and $2.5$ km/s lines.
\label{pic_ven_jup_small}}
\end{figure}
Inspired by the last subsection, we will from now on keep the density of Jupiter crossing orbits constant and focus on the combined
diffusion effects caused by Venus and the Earth. The locking of the Jupiter crossing orbits
is done in the same way as for the free orbits, see Eq.~(\ref{matris_ekv}), with the forced insertion of an identity matrix in
the time evolution operator. As mentioned above, this is justified by the fact that diffusion of Jupiter--crossing orbits is so fast that we can view these orbits as constantly being filled from the halo. We also change the interpolations between the Earth and Venus so that
Jupiter--crossing orbits are excluded (as they are filled by Jupiter).

Before going through the results, let us spend some time going through the diffusion processes in the low velocity region (as seen from the Earth).
Figure \ref{pic_ven_jup_small} is a closer view of the space of
low--velocity orbits. If we ignore the filling effects of Jupiter,
the Earth would have to diffuse WIMPs all the way from the unbound
orbits, starting at the $u=12.3$ km/s sphere. They could eventually
reach the Venus--crossing orbits to the left of the figure. Venus could
then act to diffuse the particles along \emph{its} spheres of constant
velocity. It is evident from the figure that the combined effect of
the Earth and Venus could possibly populate all orbits outside the
$u=2.5$ km/s line. By numerical simulation of the Earth-Venus system
alone, it turns out that solar depletion is so strong that
gravitational diffusion can only make a small contribution to the
particle density below $12.3$ km/s. This is no big surprise,
since comparing the scattering times in Fig.~\ref{pic_diff_tsc} and
the solar depletion times of Fig.~\ref{fig_logsoldep15}, we see 
that solar depletion is indeed very strong.

If we instead use the knowledge about the density of Jupiter crossing
orbits, the situation is very different. The Earth can
scatter particles directly from the bound Jupiter crossing orbits,
starting at $u\approx8.8$ km/s, as opposed to $12.3$ km/s for free orbits.
Furthermore the time scale of scattering, as well as the angular path the WIMPs have to
travel is much shorter, especially in the low velocity region. Hence, solar depletion will
not be as effective when we include Jupiter, as the time scales for diffusion are more comparable to the solar depletion time scales.

\begin{figure}
\center 
\epsfig{width=\columnwidth,file=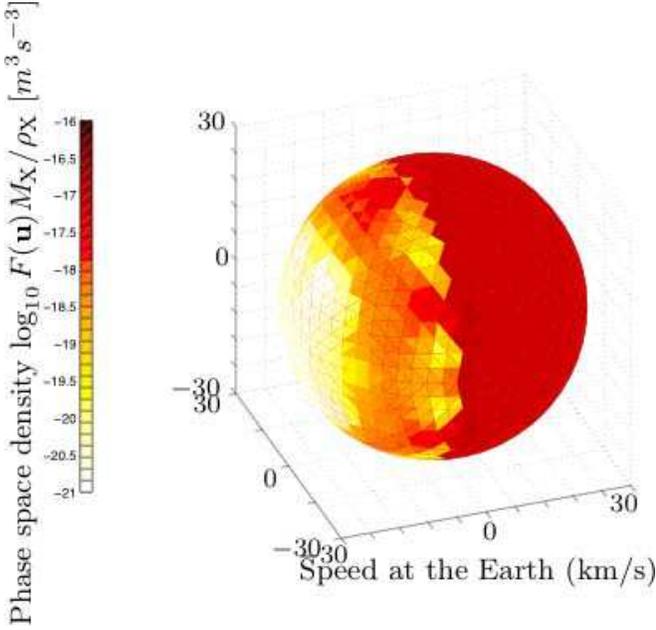}
\caption{The final phase space distribution at the $u=30$ km/s sphere.
 In understanding this figure, it may
help to take a look at the $u=30$ km/s line of figure
\ref{pic_diffprincip}, which corresponds to
the central horizontal plane of this figure. The large red region to
the right corresponds to unbound orbits. To the left (backwards 30 km/s)
the phase space density is very low, as expected from the results of
the solar depletion calculations. The leftmost part of the large red
area corresponds to Jupiter crossing orbits, which are filled with the
same density as the unbound orbits.
\label{example_spheres}}
\end{figure}

In our full calculations, we will (as mentioned above) keep Jupiter--crossing orbits fixed
and include Venus and the Earth in the diffusion process.
The calculations start with a Solar system empty of dark matter, five
billion years ago. The step size (that is, how often the diffusion
effects are added to the other planet) was  usually some hundred
thousand years. The first ten million years were typically calculated
using smaller step sizes, such as $10$ thousand years. The densities
converge to their final values within a time of $500$ million years.
An example of the resulting phase space density at a sphere of
constant velocity is given in Fig.~\ref{example_spheres}. It is
important to remember that the free distribution was averaged over a
period of $100$ million years. After such a time, the bound densities take
on their final values within about $25$\%, which is an indication that the
results might vary slightly during the galactic (half) year. In practice, this has little effect, since the typical time scales for equilibrium (see section \ref{sec:annrate}) between capture and annihilation in the Earth are much longer than that and will average out these small variations over the galactic (half) year.

\begin{figure}
\center
\epsfig{width=\columnwidth,file=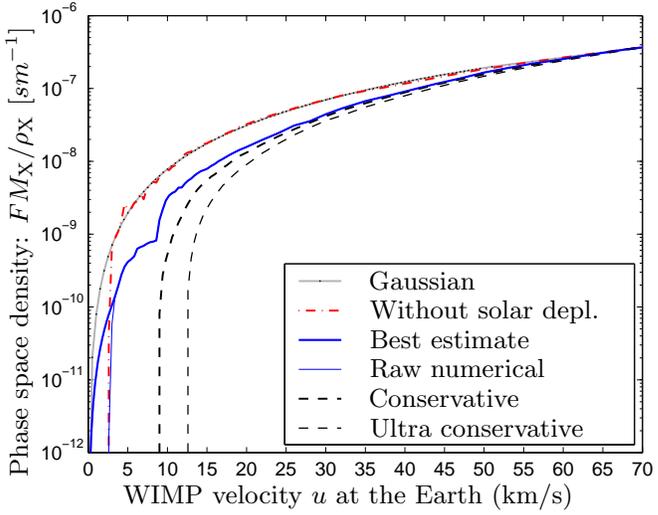}
\caption{\label{pic_the_vel_dist}
  The radial velocity distribution of Earth crossing dark matter at 
  the Earth. The curves labeled \emph{conservative} and \emph{ultra
  conservative} are the contributions from unbound, as well as unbound
  plus Jupiter crossing orbits respectively. The dash--dotted curve
  displays the result of ignoring solar depletion. 
  The blue solid line represents our best estimate, including the
  effect of the eccentricity of the Earth's orbit. The thin line is
  the raw result from the numerical routines.}
\end{figure}

The resulting velocity distributions for the slowly moving particles
are shown in Fig.~\ref{pic_the_vel_dist}. The ultra conservative
and conservative curves represent the contributions from
unbound, as well as unbound plus Jupiter--crossing orbits respectively. For these, we again see the cutoff velocities of 12.3 km/s and 8.8 km/s as explained in section \ref{sec:gravdiff}. The result of our full simulation, but ignoring solar depletion altogether is also shown. It follows the Gaussian down to about 2.5 km/s where it drops to zero. This is in perfect agreement with the results of Gould \cite{gould-diff}, and we can see this agreement as a test that our numerical routines are performing as they should. Our full numerical routines without solar depletion is a numerical implementation Gould's analytical arguments about diffusion in the solar system and our results should thus (as they do) agree in this case. 
We also show our raw numerical result, which is the outcome of our full simulation with solar depletion included. It is significantly lower than the Gaussian estimate in this low-velocity region, but not as low as the conservative (or ultra conservative) view. The general argument above that the time scales of solar depletion and diffusion are not too different and that some WIMPs should remain thus turns out to be valid. Hence, solar depletion kills some of the WIMPs at low velocities, but not as many as one could have feared. Also shown in the figure is our best estimate of the velocity distribution, which is the same as our raw numerical result, but modified at low velocities (below 2.5 km/s) to include the effect of the eccentricity of the Earth's orbit, which will be explained now.

\begin{figure}
\center
\epsfig{width=\columnwidth,file=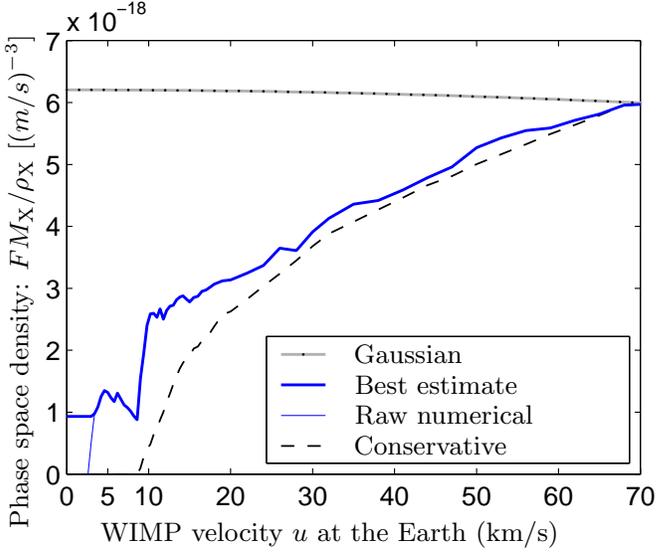}
\caption{The phase space density $F(\vect{u})$ at low velocities. 
The upper curve is the Gaussian distribution. The thin solid curve is the outcome of our
numerical simulations. The thick solid blue curve is our best estimate, where a population
at low velocities (below 2.5 km/s) has been added due to the eccentricity of the Earth's orbit (see text). The dashed line, for comparison, shows the distribution in the conservative view where only unbound plus Jupiter--crossing orbits are included.
\label{excpicture}}
\end{figure}

The diffusion effects included so far does not provide means of
filling the extremely slow ($u<2.5$ km/s) orbits. Such processes
arise when the eccentricity of the Earth's orbit is taken into
account. This could be done in a way similar to ordinary diffusion,
but since $u$ would no longer be fixed even in the one planet case,
the block diagonal one planet time evolution operator would be
polluted with new, off diagonal blocks, making the diffusion problem
much more complicated. However, the eccentricity of the Earth's orbit will mean that the Earth
diffuse slightly differently in the different parts of it's orbit. This will cause a mixing
of spheres of different $u$ and thus cause an effective diffusion in the $\hat{u}$
direction. The size of this effect can be estimated using Eq.~(2.10) of Gould's paper
\cite{gouldcosmo}. Evaluation shows that for extremely slow particle
orbits, the time scales can be as fast as one tenth of those of the
ordinary diffusion, while in most other cases they are far slower. It is
therefore quite reasonable to ignore these effects in our diffusion
treatment at higher velocities. For the $u<5$ km/s region the time scales of $u$--diffusion is
comparable to the time scales of solar depletion, which makes it
reasonable to assume that the phase space density is a slowly changing
function with respect to $u$ which means that the sharp cutoff at
$u=2.5$ km/s is not physical. To estimate the phase space density at these very low velocities, the
mean density in the $u\in[2.5,5]$ km/s region is calculated and used as a
minimum density in the whole $u\in[0,5]$ km/s region. Another approach
could have been to relocate the already existing mass to fill up the
$u<5$ km/s region evenly. However, this would underestimate the density in the
$u>2.5$ km/s region. Fig.~\ref{excpicture} compares the raw result of the
full numerical simulations, with this new best estimate and the Gaussian. The conservative view is also shown for reference.

\begin{figure}
 \center														
 \epsfig{width=\columnwidth,file=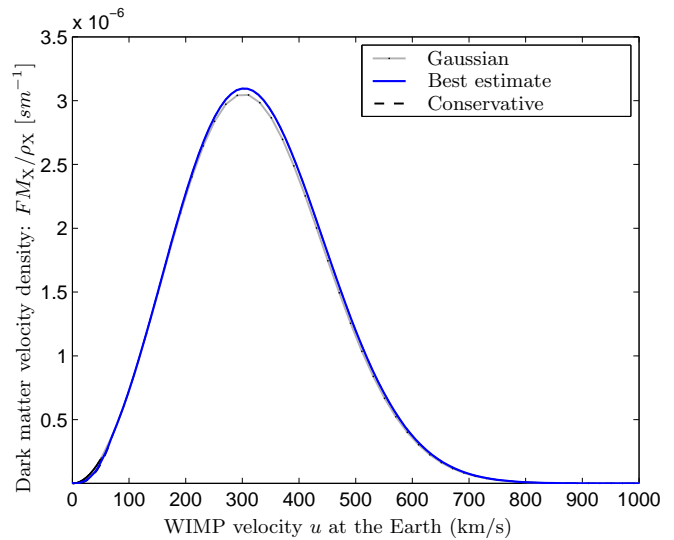}		
\caption{The radial velocity distribution of Earth crossing dark matter at 					
   the Earth, linear scale. The curves are labeled as in figure							
   \ref{pic_the_vel_dist}. Most						
   of the velocity distribution is unchanged by the considerations in						
   this report. The major difference between the Gaussian and the other						
   distributions is that latter have fewer slow particles, due to the
    effects of the solar potential.
   \label{pic_the_vel_dist_lin}}									
 \end{figure}

We have now focused on the low velocity region of the velocity distribution. In Fig.~\ref{pic_the_vel_dist_lin} we show the full velocity distribution for large velocities. At low velocities our results are confined
by the free space Gaussian and the focused free space Gaussian of the
\emph{conservative view}. Focused here means that the distribution as seen at the Earth is somewhat larger than in the halo due to the fact that particles are focused when they fall into the solar potential well. Thus, at typical galactic velocities, the
Gaussian is somewhat lower than both our best estimate and the conservative view.

\section{Capture and annihilation rates}

In the previous section, we have seen that our new estimate of the
WIMP velocity distribution is, especially at low velocities,
considerably lower than earlier estimates based on the Gaussian
approximation \cite{gould-direct}. We will here investigate how this
new velocity distribution affects first the capture rates of WIMPs in
the Earth and secondly the annihilation rates of WIMPs in the center
of the Earth. In this section, we will keep the discussion general and
in section \ref{sec:susy} we will investigate the effects for the
neutralino as a WIMP dark matter candidate.

\subsection{A new estimate of the capture rates...}
\label{sec:newcap}

\begin{figure}
\center
\includegraphics[width=\columnwidth]{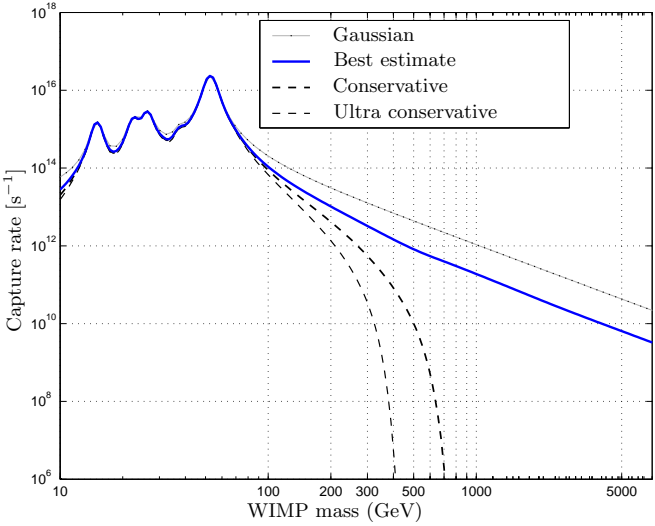}
\caption{The capture rate of dark matter. This figure shows the rate
at which dark matter particles are captured to the interior of the
Earth\label{pic_capturerate}, for a scattering cross section of
$\sigma=10^{-42}$ cm$^2$. The Gaussian--no solar depletion model gives
the highest capture.
The curves labeled \emph{ultra conservative} and \emph{conservative}
  are the contributions from unbound, as well as unbound plus Jupiter
  crossing orbits respectively. For masses above 150 GeV, our new
  capture estimate is considerably lower than that the Gaussian
  model. The peaks at low WIMP mass correspond to the masses
  of the included elements. A dark matter halo density of
  $\rho_\textsc{x}$=$0.3$~GeV$/$cm$^3$ is assumed.}
\end{figure}

Given the velocity distribution derived in the previous section, we can now calculate the capture rate in the Earth with this velocity distribution. We will use the full expressions for the capture rate as derived by Gould in \cite{gould-resonant}, but will also compare with the usual Gaussian approximation (as derived in \cite{gould-diff}), as that is what most people use to calculate the capture rates.

The calculation of the capture rates for an arbitrary velocity distribution is given in \cite{gould-resonant}, we will here only briefly outline how the calculation is done.

We divide the Earth into shells, where the capture from element $i$ in each shell (per unit shell volume) is given by \cite{gould-resonant}[Eq.~(2.8)]
\begin{equation}
\label{eq:dcdv}
    \frac{dC_i}{dV} = \int_0^{u_{max}} du \frac{\tilde{f}(u)}{u} w \Omega_{v,i}^-(w)
\end{equation}
where $\tilde{f}(u)$ is the velocity distribution (normalized such that $\int_0^\infty \tilde{f}(u) = n_\chi$ where $n_\chi$ is the number density of WIMPs\footnote{We introduce the velocity distribution $\tilde{f}(u)$ here to be as close as possible to Gould's expressions. This distribution is related to $F(u)$ in section \ref{sec:galhalo} through $\tilde{f}(u)=4\pi u^2 F_u (u)$.}. The expression $\Omega_{v,i}^-(w)$ is related to the probability that we scatter to orbits below the escape velocity. $w$ is the velocity at the given shell and it is related to the velocity at infinity $u$ and the escape velocity $v$ by $w = \sqrt{u^2 + v^2}$.
The upper limit of integration is a priori set to $u_{max} = \infty$, but we will see below that due to kinematical reasons we can set it to a lower value (Eq.~(\ref{eq:umax}) below).
If we allow for a form factor suppression of the form \cite{gould-resonant}[Eq.~(A3)]
\begin{equation}
   |F(q^2)|^2 = \exp\left( - \frac{\Delta E}{E_0} \right)
\end{equation}
with \cite{gould-resonant}[Eq.~(A4)]
\begin{equation}
\label{eq:e0}
   E_0 = \frac{3 \hbar^2}{2m_\chi R^2}
\end{equation}
we can evaluate $w \Omega_{v,i}^-(w)$ and arrive at the expression \cite{gould-resonant}[Eq.~(A6)]
\begin{eqnarray}
\label{eq:womega}
   w \Omega_{v,i}^- (w) & = & \sigma_i n_i \frac{\mu_+^2}{\mu}2 E_0 \left[
   e^{- \frac{m_\chi u^2}{2E_0}} - e^{-\frac{\mu}{\mu_+^2}m_\chi \frac{u^2+v^2}{2E_0}}
   \right] \nonumber \\
   & & \Theta\left( \frac{\mu}{\mu_+^2} - \frac{u^2}{u^2+v^2} \right)
\end{eqnarray}
where we have introduced
\begin{equation}
   \mu = \frac{m_\chi}{m_i} \quad ; \quad \mu_\pm = \frac{\mu \pm 1}{2}
\end{equation}
with $m_i$ the mass of element $i$. The Heaviside step function $\Theta$ plays the role of only including WIMPs that can scatter to a velocity lower then the escape velocity $v$. To simplify our calculations we can drop this step function in Eq.~(\ref{eq:womega}) and instead set the upper limit of integration in Eq.~(\ref{eq:dcdv}) to
\begin{equation}
\label{eq:umax}
  u_{max} = \sqrt{\frac{\mu}{\mu_-^2}} v
\end{equation}
We also need the scattering cross section on element $i$, which can be written as
\cite{susydm}[Eq.~(9-25)]
\begin{equation}
\label{eq:sigma_i}
   \sigma_i = \sigma_p A_i^2 \frac{(m_\chi m_i)^2}{(m_\chi+m_i)^2} 
   \frac{(m_\chi + m_p)^2}{(m_\chi m_p)^2}
\end{equation}
where $A_i$ is the atomic number of the element, $m_p$ is the proton mass and $\sigma_p$ is the scattering cross section on protons.

We now have what we need to calculate the capture rate. In Eq.~(\ref{eq:dcdv}) we integrate over the velocity for our chosen velocity distribution. We then integrate this equation over the radius of the Earth and sum over all the different elements in the Earth,
\begin{equation}
\label{eq:cfinal}
   C = \int_0^{R_\oplus} dr \sum_i \frac{dC_i}{dV} 4 \pi r^2
\end{equation}

\begin{table}
  \centering 
  \begin{tabular}{lccc} \hline
& Atomic & \multicolumn{2}{c}{Mass fraction} \\ \cline{3-4} 
 Element & number & Core & Mantle \\ \hline
  Oxygen, O     & 16 & 0.0   & 0.440   \\
  Silicon, Si   & 28 & 0.06  & 0.210   \\
  Magnesium, Mg & 24 & 0.0   & 0.228   \\
  Iron, Fe      & 56 & 0.855 & 0.0626  \\
  Calcium, Ca   & 40 & 0.0   & 0.0253  \\
  Phosphor, P   & 30 & 0.002 & 0.00009 \\
  Sodium, Na    & 23 & 0.0   & 0.0027  \\
  Sulphur, S    & 32 & 0.019 & 0.00025 \\
  Nickel, Ni    & 59 & 0.052 & 0.00196 \\
  Aluminum, Al  & 27 & 0.0   & 0.0235  \\
  Chromium, Cr  & 52 & 0.009 & 0.0026  \\ \hline
\end{tabular}
  \caption{The composition of the Earth's core and mantle. The core mass fractions are from \cite{earthcomp}[Table 4] and the mantle mass fractions are from \cite{earthcomp}[Table 2].} \label{tab:earthcomp}
\end{table}

The capture rates depend on the mass and distribution of the elements in
the Earth. The most important elements are iron, silicon, magnesium
and oxygen, of which iron is by far most important for WIMP masses over
$100$ GeV. We use the Earth density profile as given in \cite{EncBrit} and for the element distribution within the Earth we use the values given in \cite{earthcomp}[Table 2 for the mantle and Table 4 for the core]. These values are listed in Table~\ref{tab:earthcomp}.

Fig.~\ref{pic_capturerate} shows the calculated capture rates, to be
compared with that of a Gaussian distribution, with the Earth in free
space. The Gaussian distribution is the one of equation
(\ref{solgauss}) in section \ref{sec:galhalo}. This common model is
equivalent to having the Earth taking the place of the Sun and
removing the solar system (this is how weak capture into the Sun is
usually calculated). The scattering cross section between the nucleons
and the WIMPs determines the normalization only, and was taken to be
$10^{-42}$ cm$^2$ in Fig.~\ref{pic_capturerate}. 
We also show the resulting capture rates in the conservative and ultra conservative view, where the cutoffs at about 710 GeV and 410 GeV are clearly seen. These cutoff masses are higher than those in Gould and Alam \cite{gould-conserv} as we have used the full integration over the Earth and not the average properties as in \cite{gould-conserv} (see section \ref{sec:gravdiff} for a discussion of these cutoff masses). 

\begin{figure}
\center
\includegraphics[width=\columnwidth]{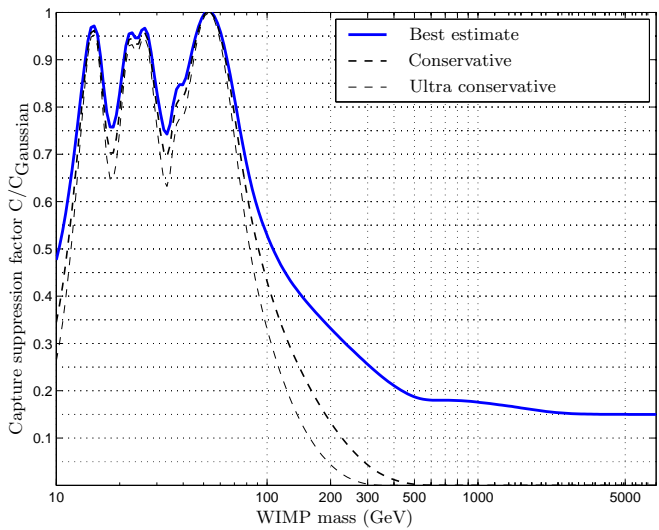}
\caption{The ratio between the capture in various models and
that of a Gaussian distribution in free space.\label{rategauss} The
figure displays the quotient of the weak WIMP capture rates in the
Earth in various models, and the capture given in the case of the
commonly used Gaussian distribution. 
\label{pic_major_results}}
\end{figure}

It is of course interesting to compare the calculated capture rate
with that given by the commonly used Gaussian distribution. This is
done in Fig.~\ref{rategauss}, where we divide by the capture rate in the Gaussian approximation. We clearly see that below ~100 GeV, the different calculations agree to within about a factor of two. At higher masses the suppression is almost an order of magnitude, but not as bad as the feared conservative or ultra conservative views.

\subsection{...and the annihilation rates}
\label{sec:annrate}

We have seen that our new estimate of the capture rate in the Earth is, especially at higher masses, considerably lower than the usual estimate based on the Gaussian approximation \cite{gould-diff}. Since the neutrino-induced muon rates do not directly depend on the capture rate, but instead on the annihilation rate, we will here investigate how the annihilation rates are affected.

The evolution equation for 
the number of WIMPs, $N$, in the Earth is given by
\begin{equation} \label{eq:Nevol}
  \frac{dN}{dt} = C - C_{A} N^2 - C_{E} N
\end{equation}
where the first term is the WIMP capture, the second term 
is twice the annihilation rate $\Gamma_{A} = 
\frac{1}{2} C_{A} N^2$ and the last term is WIMP evaporation. 
The evaporation term can be neglected for WIMPs heavier than 
about 5--10 GeV \cite{gould-resonant} and 
since we are not interested in these low-mass 
WIMPs we can safely drop the last term in Eq.~(\ref{eq:Nevol}).
If we solve Eq.~(\ref{eq:Nevol}) for the annihilation rate 
$\Gamma_{A}$ we get
\begin{equation} \label{eq:GammaA}
  \Gamma_{A} = \frac{1}{2} C \tanh^2 \frac{t}{\tau} , \qquad \tau = 
  \frac{1}{\sqrt{C C_{A}}}
\end{equation}
where $\tau$ is the time scale for capture and annihilation 
equilibrium to occur.  In the Sun, equilibrium will for many WIMP models
have occurred and the annihilation rate is at `full strength', $\Gamma_{A} 
\simeq \frac{1}{2} C$. In this case the annihilation rate is directly proportional
to the capture rate. However, in the Earth, equilibrium has often not occurred, and we will have the more complex relation between capture and annihilation rate, Eq.~(\ref{eq:GammaA}). In the next section, we will show this for an explicit example, the neutralino in the Minimal Supersymmetric Standard Model (MSSM). Before looking at specific MSSM models, let's analyze Eq.~(\ref{eq:GammaA}) to see the general trends. Let's denote the capture and annihilation rates in the usual Gaussian approximation by $C^G$ and $\Gamma_A^G$ respectively, whereas our new estimates are denoted $C$ and $\Gamma_A$. Using the fact that the constant $C_A$ is the same in both scenarios, we can then write
\begin{equation}
  \frac{\Gamma_A}{\Gamma_A^G} = \frac{C}{C^G}   
  \frac{\tanh^2\left(\sqrt{\frac{C}{C^G}}\frac{t_\odot}{\tau}\right)}
  {\tanh^2\left(\frac{t_\odot}{\tau}\right)}
  \simeq \left\{
  \begin{array}{lcl}
  \frac{C}{C^G} & ; & t_\odot \gg \tau \\[1ex]
  \left(\frac{C}{C^G}\right)^2 & ; & t_\odot \ll \tau
  \end{array} \right.
  \label{eq:annsup}  
\end{equation}
Hence, if equilibrium has occurred, the annihilation rate (and thus the neutrino-induced muon fluxes) are suppressed with the same factor as the capture rates, but if equilibrium has not occurred, the annihilation rate is suppressed with the square of the capture rate suppression factor, i.e.\ the suppression is amplified.

\section{Application to the supersymmetric neutralino}
\label{sec:susy}

So far, we have discussed the effects of our new estimate of the
velocity distribution in general terms. We have seen that our estimate
of the velocity distribution is significantly different from previous
estimates at low velocities. We have also seen that the capture rates,
especially at higher WIMP masses are significantly reduced with a
factor $C/C^G$. Hence, the annihilation rates (and
the expected neutrino-induced muon fluxes) are reduced by a
factor that lies in between $(C/C^G)^2$ and $C/C^G$. We now want to
investigate this suppression factor further and analyze the
effects on the neutrino-induced muon fluxes. For this we need an
explicit WIMP candidate. We will here assume that the WIMP is the
lightest neutralino, that arises as a natural dark matter candidate in
supersymmetric extensions of the standard model. In the next
subsection, we will briefly go through the supersymmetric model we
work in and will then continue to investigate the effects of our new
velocity distribution on the annihilation rates and the
neutrino-induced muon fluxes.

\subsection{The neutralino as a dark matter candidate}

We will assume that the WIMP is the lightest neutralino in the Minimal Supersymmetric Standard Model (MSSM), i.e.\ the lightest neutralino, 
$\tilde{\chi}^0_1$, is defined as the lightest mass eigenstate obtained 
from the superposition of four spin-1/2 fields, the Bino and Wino gauge 
fields, $\tilde{B}$ and $\tilde{W}^3$, and two neutral CP-even Higgsinos,
$\tilde{H}^0_1$ and $\tilde{H}^0_2$:
\begin{equation}
  \tilde{\chi}^0_1 = 
  N_{11} \tilde{B} + N_{12} \tilde{W}^3 + 
  N_{13} \tilde{H}^0_1 + N_{14} \tilde{H}^0_2\;.
\end{equation}
For a recent review of the MSSM and the neutralino as a dark matter candidate, see \cite{lbreview}. The parameters of our phenomenologically inspired MSSM model are the Higgsino mass parameter $\mu$, the gaugino mass parameter $M_2$, the ratio of the Higgs vacuum expectation values $\tan\beta$, the sfermion mass scale $M_{\tilde{q}}$, the mass of the CP-odd Higgs boson $m_A$, and the trilinear couplings for the third generation squarks $A_t$ and $A_b$. We have made extensive scans of these parameters and have currently about a couple of hundred thousand models in our model database. 

For our actual calculations we use the DarkSUSY package~\cite{darksusy}. We only select those models that do not violate present accelerator bounds. The neutralino naturally  has a relic density in the right ballpark, and we will further restrict this range by selecting only models with a relic density in the range $0.05 \le \Omega_\chi h^2 < 0.2$. This range is a bit larger than the current best estimates \cite{wmap}, but to be conservative we choose work with this larger range. When calculating the relic density, we have included coannihilations between neutralinos and charginos (coannihilations also with sfermions in the MSSM is the subject of a future publication).

\subsection{Neutralino capture and annihilation}

\begin{figure}
\centerline{\epsfig{file=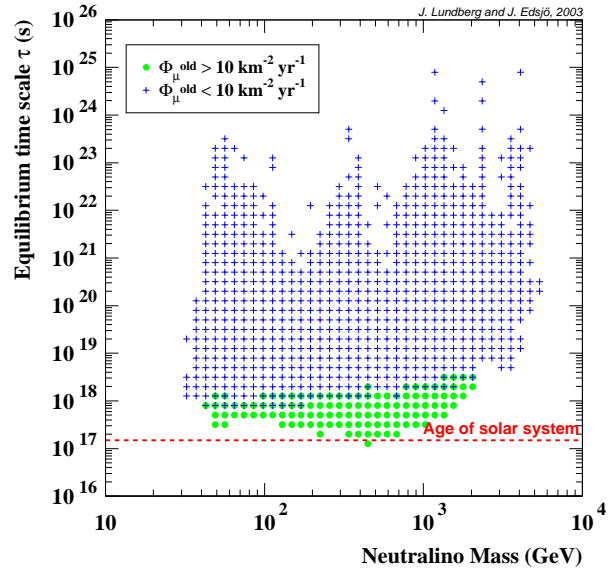,width=\columnwidth}}
\caption{The equilibrium time scales $\tau$ for a set of MSSM models. As seen, all our models have equilibrium time scales longer than the age of the solar system. The green dots are models for which the neutrino-induced muon fluxes (with the usual Gaussian approximation) are larger than 10 km$^{-2}$ yr$^{-1}$ and the blue plus signs indicate models with smaller fluxes.}
\label{fig:tau_earth}
\end{figure}

We will here investigate how the annihilation rates are affected for specific MSSM models. In Fig.~\ref{fig:tau_earth} we show typical equilibrium time scales, $\tau$, for a set of MSSM models. As seen, the typical equilibrium time scales are much longer than the age of the solar system, $t_\odot \simeq 4.5 \cdot 10^9$ years, and hence equilibrium has often not occurred in the Earth. 

\begin{figure}
\centerline{\epsfig{file=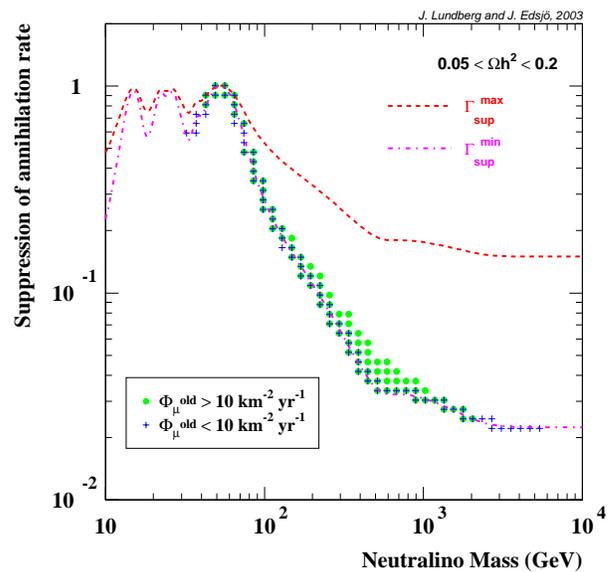,width=\columnwidth}}
\caption{$\Gamma_A/\Gamma_A^G$ versus the neutralino mass $m_\chi$. The limiting cases for $t_\odot \gg \tau$ and $t_\odot \ll \tau$ are indicated in the figure. Most models have annihilation rate suppressions close to the lower curve since equilibrium has most often not occurred in the Earth.}
\label{fig:annsup-mx}
\end{figure}

As equilibrium has not occurred in the Earth, we can use Eq.~(\ref{eq:GammaA}) to see how the decrease in $C$ will affect $\Gamma_A$. 

In Fig.~\ref{fig:annsup-mx} we show, for a set of MSSM models, how the annihilation rates are decreased. We also show the limiting cases for $t_\odot \gg \tau$ and $t_\odot \ll \tau$. We can clearly see that for most models, as equilibrium has not occurred, we are close to the $(C/C^G)^2$ suppression of the annihilation rates.

\subsection{Neutrino-induced muon fluxes from the Earth}

\begin{figure*}
\centerline{\epsfig{file=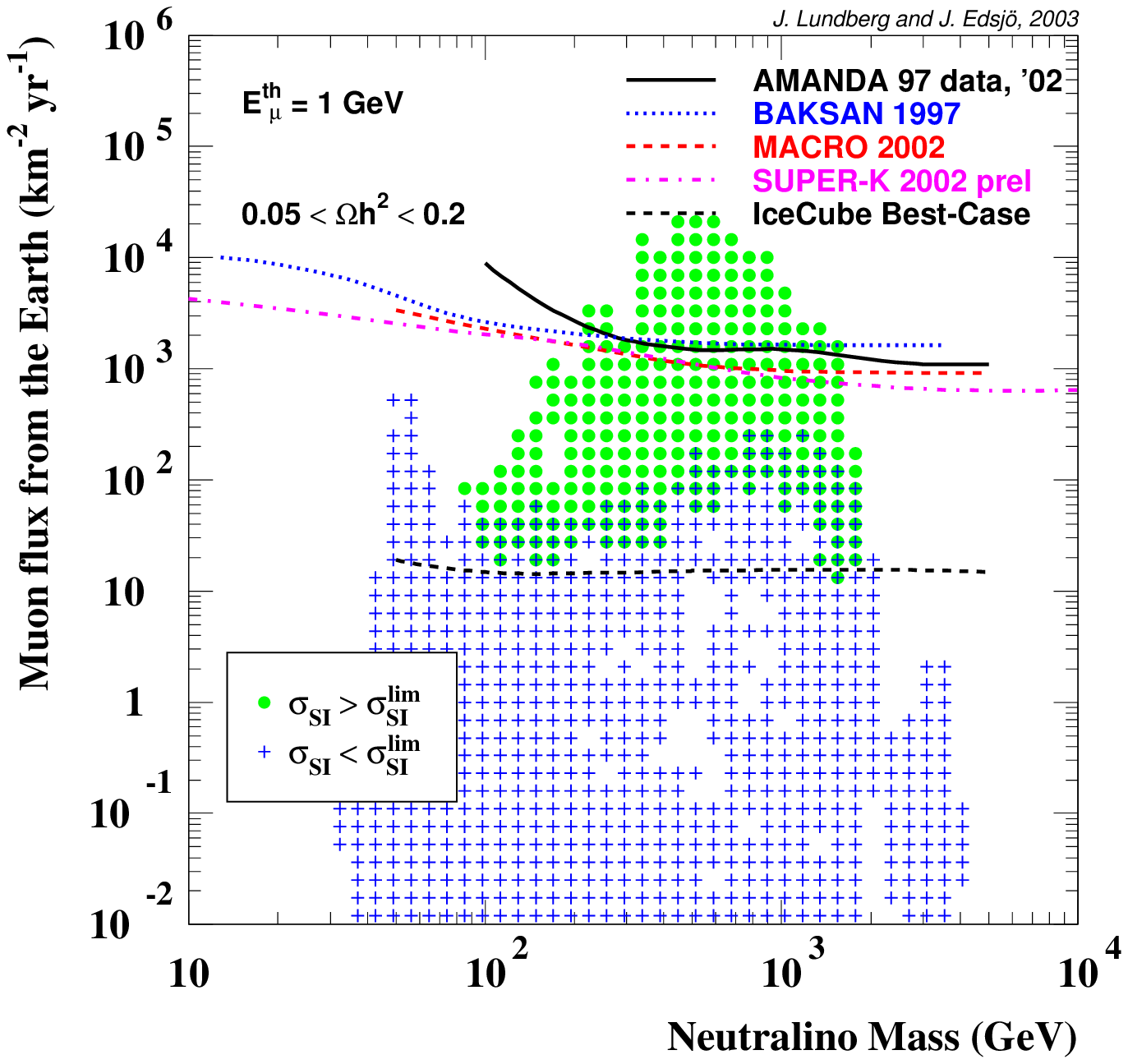,width=0.49\textwidth}
\epsfig{file=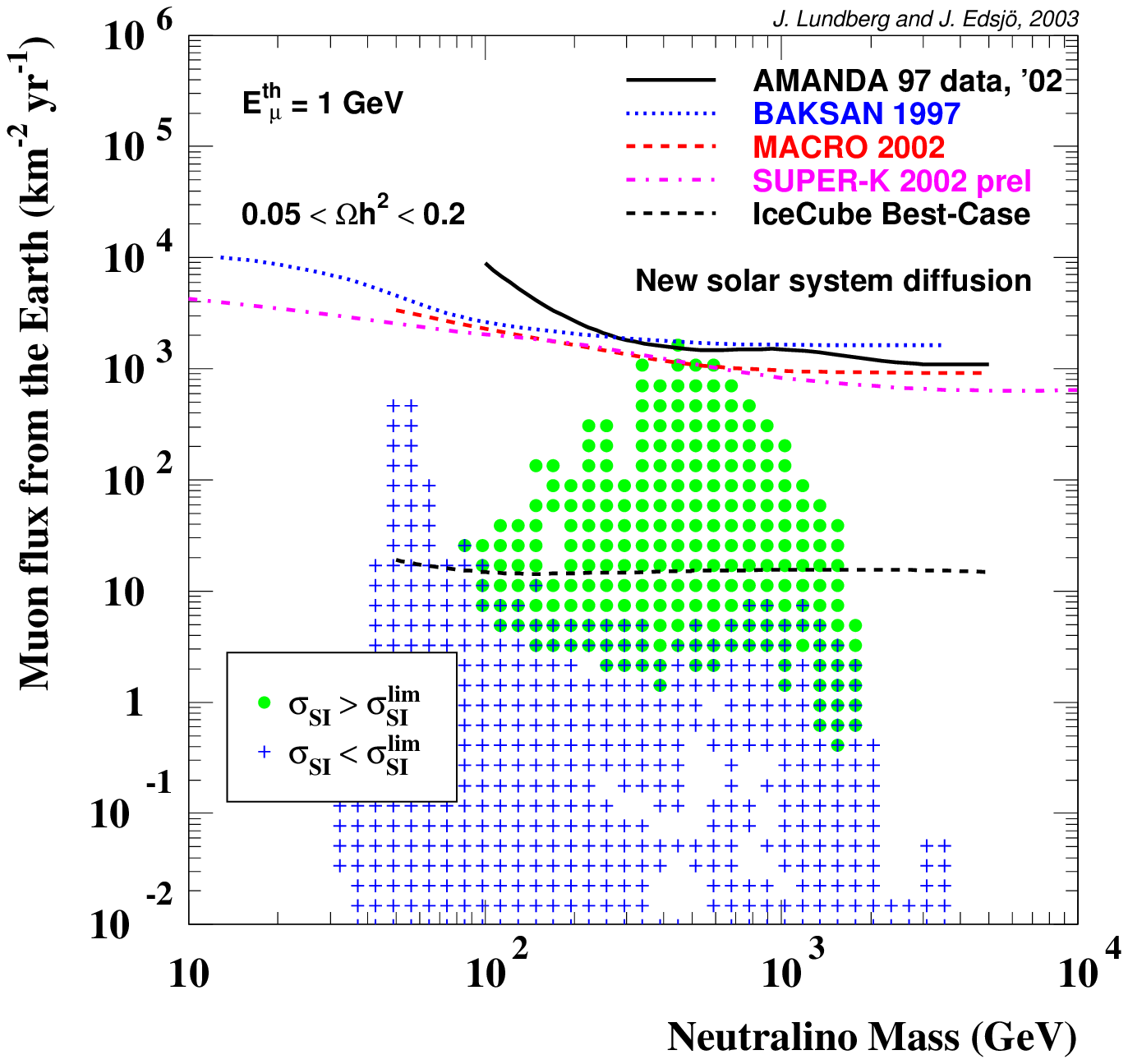,width=0.49\textwidth}}
\caption{In the left panel we show the neutrino-induced muon fluxes in the
standard Gaussian approximation, whereas in the right panel we show the fluxes based on our new estimate of the WIMP diffusion in the solar system. We also show the current limits of a few neutrino telescopes and an optimistic estimate for the future IceCube sensitivity. The current direct detection limit by the Edelweiss experiment \cite{edelweiss-2002} is also shown. Models that are excluded by Edelweiss are indicated by green circles, whereas models that are not excluded are indicated with blue crosses.}
\label{fig:rea1-mx}
\end{figure*}

So, given our calculated suppression of the annihilation rates,
the neutrino-induced muon fluxes will also be suppressed by the same amount. We now ask ourselves if this suppression is too big to make the neutrino-induced muon fluxes too low to be observable in the MSSM. In Fig.~\ref{fig:rea1-mx} we show in the left panel the neutrino-induced muon fluxes with the old Gaussian approximation. In the right panel, we show the neutrino-induced fluxes with our new estimate of the WIMP velocity distribution. We also indicate current limits from neutrino telescopes (Baksan \cite{baksan-wimp}, Macro \cite{macro-wimp}, Amanda \cite{amanda-wimp} and Super-Kamiokande \cite{superk-wimp}) and anticipated sensitivities for 
future neutrino telescopes like IceCube \cite{icecube-wimp}. Note that the IceCube limit shown here is a probably too optimistic, but we show it as limiting case beyond which a 1 km$^3$ neutrino telescope will not reach.
For comparison, we also indicate the current direct detection limit by the Edelweiss experiment \cite{edelweiss-2002}. Models that are excluded by Edelweiss are indicated by green circles, whereas models that are not excluded are indicated with blue crosses.

Comparing the left and the right figure,  
we clearly see that there is a significant suppression of the rates above about 100 GeV and above about 2000 GeV, the fluxes are too low to be observable even with future detectors.
In the range between 100 GeV and 2000 GeV, where future neutrino telescopes still have a chance to detect a signal from the Earth, the prospects for doing so is clearly diminished with our new estimate of the fluxes. Especially if one considers that all of the observable models in that range are already excluded by direct detection experiments. Note, however, that the comparison between direct detection and neutrino telescopes that we have done here is for a Maxwell-Boltzmann velocity distribution. As direct detection experiments are primarily sensitive to the high velocity tail of the distribution, whereas neutrino telescopes are sensitive to the low velocity tail, the correlation between the two signals need not be as large as indicated in Fig.~\ref{fig:rea1-mx} for a more realistic distribution. Below 100 GeV, the neutrino signal from the Earth is not reduced much with our new velocity distribution. In this range, neutrino telescopes are also in general more sensitive than direct detection experiments.

\section{Conclusions}
\label{sec:conclusions}

We have made a new estimate of the velocity distribution of WIMPs at the Earth due to diffusion in the solar system. We have included gravitational diffusion due to the Earth, Venus and Jupiter and depletion due to solar capture. Compared to the standard approximation (i.e.\ that the solar diffusion can be approximated by the Earth being in free space and seeing the unperturbed Gaussian halo velocity distribution), our estimate is significantly lower at low velocities (below about 70 km/s). The main reason for this is that solar capture diminishes the WIMP population at these low velocities. If it were not for solar capture, our results would confirm the results of Gould \cite{gould-diff}, i.e.\ that the velocity distribution as seen at the Earth is close to that we would see if the Earth was in free space. The diffusion effects of Jupiter, Earth and Venus would make the distribution look Gaussian, apart from a hole in the distribution below 2.5 km/s. This hole would however be filled due to the eccentricity of the Earth's orbit. However, solar capture suppresses the velocity distribution by about an order of magnitude at low velocities and this suppression propagates into a suppression of the same order of magnitude in the capture rate. 

Since the annihilation rates depend on the capture rates, the
annihilation rates are also suppressed. The amount of suppression, however, depends on if capture and annihilation is in equilibrium or not. If it is in equilibrium the annihilation rate suppression is the same as the capture rate suppression, but if we are far from equilibrium, the annihilation rate suppression is equal to the capture rate suppression squared. 

For one of the prime WIMP dark matter candidates, the neutralino in the Minimal Supersymmetric Standard Model (MSSM),  it turns out that these are typically not in equilibrium and thus the annihilation rate suppression is equal to the capture rate suppression squared.
The net result is that the annihilation rates will start being suppressed above about 100 GeV, and reaches a maximal suppression of about $10^{-2}$ at around 1 TeV. Above about 2 TeV, the expected fluxes are so low that future neutrino telescopes will not have enough sensitivity to see these.

Finally, a word of caution should be applied to the interpretation of these new results. Even if we have done what we can to make sure that our new estimate is correct, there are still approximations done and numerical uncertainties that need to be considered. E.g., in principle one would like to do a full numerical simulation of the full diffusion process with an arbitrary halo distribution as input. That is not numerically feasible to do so instead we have relied on numerical simulations for the solar capture and on analytical calculations and arguments for the diffusion process. These analytical calculations are approximations with the aim to describe the diffusion processes correct in average. We think that these approximations are reasonable, but one should keep in mind that there are uncertainties involved in these approximations. At higher masses, above about 1 TeV, we are very sensitive to the very details of the velocity distribution at very low velocities (a few km/s). We have assumed that the eccentricity of the Earth's orbit fills the hole below 2.5 km/s. If this would not be the case, the suppression for high masses would be even larger than depicted here.

\section*{Acknowledgments}
J.~Edsj\"o thanks the Swedish Research Council for support.

\appendix

\section{Comparison with Farinella's calculation of near Earth asteroids}
\label{app:farinella}

\begin{table}[htb!]
\center
\caption{The asteroids integrated by Farinella et al.
The numbers given are the times at which the asteroid collided with the
Sun, or was ejected, in thousands of years. The (**) marks asteroids we have not
calculated. The following asteroids survived the full
two million year period, according to Farinella et al.\ and our calculation (and are not included in the table): 
1972 RB, 1981 QB, 1981 QN1, 1982 TA , 1984 KB**, 1990 OA, 1990 SM,
1991 EE , 1991 VC, 1992 EU , 1992 RD, 1992 SY, 1992 SZ, 1998 CC1,
1998 PA, Beltrovata, Dionysus , Dorchester, Grieve, Hiltner, Krok,
Oljato, Poseidon, Taurinensis, Verbano, Verenia , Wisdom, Zeus.
\label{astrotable} }
\begin{tabular}{l  | r r c @{} | r r  }
   \multicolumn{3}{c}{\hspace*{7em}Mercury pack}  & \multicolumn{3}{c}{\hspace*{2em}Farinella et al.} \\
\hline
1971 SC         &            &  **	&&sun	&1400	\\
1983 LC		&  	sun   &   42   	&&sun	&810	\\ 
1988 NE		&   	sun   &   1062 	&&sun	&950	\\
1988 VP$_4$     &            &  **	&&sun	&1470	\\
1989 DA	        &	sun   &   369 	&& 	&	\\
1990 HA	     	& sun   &  1985 	&&eject	&450	\\
1990 TG$_1$	 & 	eject &  362 	&&eject	&420	\\
1990 TR	         &   	eject & 1449 	&& 	&	\\
 1991 AQ	 &	sun   &   456 	&& 	&	\\
 1991 BA	 &            &        	&&sun	&120	\\
 1991 GO	 &    	      &  	&&sun	&600	\\
 1991 SZ           &         &  	&&sun	&1860	\\
 1991 TB$_2$	 &      sun   &   625 	&&sun	&30	\\
 1991 VP$_5$     &         &  **	&&sun	&570 	\\
 1992 SY	 &       sun  &  1509 	&& 	&	\\	
 6344 P-L	 &	eject &  362 	&& 	&	\\
 Adonis		 &  	sun   &   1214 	&&sun	&900	\\
 Cuno		 &    	sun   &   1274 	&&eject	&640	\\
 Encke		 &            &  	&&sun	&90	\\
 Hephaistos      &	sun   &   143	&&sun	&110	\\
 Mithra		 &	sun   &   205 	&&sun	&180	\\
 Ojato		 & 	sun   &   328 	&&sun	&360 	\\
 Toutatis	 & 	eject & 79 	&&eject	&640 	\\
\end{tabular}
\end{table}

This appendix considers the asteroids which fates were investigated
by Farinella et al.\ \cite{farinella}. They calculated the fates
of about 47 asteroids of which most are near Earth asteroids (NEAs).
The result was that about a third of the considered asteroids were
ejected to hyperbolic orbits or driven into the sun in less than 2
million years. This led Gould to consider the possibility that the
population of gravitationally bound dark matter is heavily reduced by
solar capture. This in turn lead to the Conservative and Ultra
conservative views discussed in the introductory sections.

As a test of the \textsf{Mercury} integration package \cite{merc}, we have
repeated the calculations of Farinella et al.\ using both
\textsf{Bulirsch-Stoer} \cite{bulstoer} and fifteenth--order \textsf{Radao} \cite{radao}.
These are quite complicated methods specially developed for solving
the many-body problem.

The actual fates of specific asteroids is of course dependent of the
method used, and the accuracy parameters of the calculation. Even with
very high accuracy, convergence can not be expected since numerical
errors propagate exponentially in chaotic systems. The initial
conditions of our calculations are those of the online asteroid database
at U.S.\ Naval Observatory, epoch 11-22-2002. In addition to the time
passed, some asteroids have been observed many times since 1994.
However, one can still hope to imitate the general behavior by looking
at a large set of initial values, regardless of what they represent;
asteroids or WIMPs.

The results for the \textsf{Bulirsh-Stoer} method are presented in table \ref{astrotable}. Of the 47 objects,
four were ejected from the solar system and twelve were captured by
the Sun in our calculations, whereas four were ejected and 14 were captured by the Sun in Farinella et al.'s calculations. We cannot expect to get exactly the same results on an individual basis, but are satisfied to see that we get roughly the same behavior as
Farinella et al.



\begin{thebibliography}{99}

\bibitem{wmap}
C.L.~Bennett et al., 
\emph{First Year Wilkinson Microwave Anisotropy Probe (WMAP) Observations: Preliminary Maps and Basic Results},
Astrophys.\ J.\ Suppl.\ {\bfseries 148} (2003) 1 [astro-ph/0302207].
\emph{First Year Wilkinson Microwave Anisotropy Probe (WMAP) Observations: Preliminary Maps and Basic Results}

\bibitem{earth-nus}
K.~Freese, Phys.\ Lett.\ {\bf B167} (1986) 295;
L.~Krauss, M.~Srednicki and F.~Wilczek, Phys.\ Rev.\ {\bf D33} (1986) 
2079;
T.~Gaisser, G.~Steigman and S.~Tilav, Phys.\ Rev.\ {\bf D34} (1986) 
2206.

\bibitem{press-spergel}
W.H.~Press and D.N.~Spergel, 
Astrophys.\ J.\ {\bfseries 296}(1985) 679.

\bibitem{gould-resonant}
A.~Gould,
\emph{Resonant enhancements in weakly interacting massive particle capture by the
  Earth},
Astrophys.\ J.\ {\bfseries 321} (1987) 571.

\bibitem{gould-direct}
A.~Gould, 
\emph{Direct and indirect capture of weakly interacting massive particles by
the Earth},
Astrophys.\ J.\ {\bfseries 328}(1988) 919.

\bibitem{gould-diff}
A.~Gould, 
\emph{Gravitational diffusion of solar system WIMPs},
Astrophys.\ J.\ {\bfseries 368} (1991) 610.

\bibitem{farinella}
P.~Farinella, C.~Froeschl\'e,
  C.~Froeschl\'e, R.~Gonczi, G.~Hahn, A.~Morbidelli and G.B.~Velsecchi,
  \emph{Asteroids falling into the Sun},
  Nature {\bfseries 371} (1994) 314).

\bibitem{gould-conserv}
A.~Gould and S.M.K~Alam,
\emph{Can heavy WIMPs be captured by the Earth?},
Astrophys.\ J.\ {\bfseries 549} (2001) 72.

\bibitem{opik}
{\"O}pik, 
\emph{Interplanetary Encounters: Close-Range Gravitational Interaction}, 
Elsevier Scientific Publishing company 1976.

\bibitem{gladman}
B.~Gladman et al., Science {\bfseries 277} (1997) 197.

\bibitem{migliorini}
F.~Migliorini et al., Science {\bfseries 281} (1998) 2022.

\bibitem{susydm}
G.~Jungman, M.~Kamionkowski, K.~Griest, 
  \emph{Supersymmetric dark matter},
  Phys.\ Rep. {\bfseries 267} (1996) 195.

\bibitem{scheck}
F.~Scheck, 
\emph{Mechanics: From Newton's Laws to Deterministic Chaos}, 3rd Edition,
Springer-Verlag 1999.

\bibitem{merc}
J.E.~Chambers, 
\emph{A Hybrid Symplectic Integrator that Permits Close Encounters between Massive Bodies},
Mon.\ Not.\ R.\ Astron.\ Soc.\ {\bfseries 304} (1998) 793.

\bibitem{radao}
E.~Everhart, \emph{An efficient integrator that uses Gauss-Radao Spacing}, 
proceedings of the International Astronomical Union, Rome, Italy, 11--15 June 1984, \emph{Dynamics of Comets: Their Origin and Evolution}, edited by 
A.~Carusi and G.B.~Valsecchi.
  
\bibitem{bulstoer}
J.~Stoer and R.~Bulirsch, in W.H.~Press et al.\ 1992,
  \emph{''Numerical Recipes in Fortran''},
  Cambridge Univ. Press, 1992.

\bibitem{wisdom}
J.~Wisdom and M.~Holman,
\emph{Symplectic maps for the n-body problem},
Astroph.\ J.\ {\bfseries 102} (1991) 1528;
J.~Wisdom, M.~Holman and J.~Touma,
\emph{Symplectic Correctors},
Fields Institute Communications {\bfseries 10} (1996) 217.


\bibitem{migrjup}
S.I.~Ipatov and J.C.~Mather, 
  \emph{Migration of the Jupiter-family comets and resonant asteroids to
  near-Earth space},
astro-ph/0303219.

\bibitem{Jupitermass}
T.~Guillot,
 \emph{A comparison of the interiors of Jupiter and Saturn},
Plan.\ Space Sci.\ {\bfseries 47} (1999) 1183.

\bibitem{EncBrit}
\emph{The Earth: its properties, composition, and structure},
Britannica CD, Version 99 \copyright 1994--1999.
Encyclop{\ae}dia Britannica, Inc.

\bibitem{gouldcosmo}
A.~Gould, 
\emph{Cosmological density of WIMPs from solar and terrestrial
annihilations},
Astrophys.\ J.\ {\bfseries 388} (1992) 338.

\bibitem{earthcomp}
W.F.~Mcdonough, Treatise on Geochemistry, Vol 2, Elsevier, 2003. (The values for the Earth composition are very close to those in The Encyclopedia of
Geochemistry, Eds. Marshall and Fairbridge, Klower Acadmic Publ., 1998.)

\bibitem{lbreview}
L.~Bergstr\"om,
\emph{Nonbaryonic dark matter: observational evidence and detection methods},
Rep.\ Prog.\ Phys.\ {\bfseries 63} (2000) 793.

\bibitem{darksusy}
P.~Gondolo, J.~Edsj\"o,
  P.~Ullio, L.~Bergstr\"om, M.~Schelke and E.A.~ Baltz, 
\emph{DarkSUSY - A numerical package for supersymmetric dark matter calculations},
astro-ph/0211238.

\bibitem{edelweiss-2002}
A.~Benoit et al., Phys.\ Lett.\ {\bfseries B545} (2002) 43.

\bibitem{baksan-wimp}
M. Boliev {\em et al.}. {\it Proceedings of Dark Matter in
Astro and Particle Physics, 1997}. H.~V.~Klapdor-Kleingrothaus and
Y.~Ramachers, eds., (Worl Scientific, Singapore, 1997), p. 711. See
also O. Suvorova, hep-ph/9911415.

\bibitem{macro-wimp}
M.~Ambrosio {\em et al.} Phys. Rev. {\bfseries D60}, 082002 (1999).

\bibitem{amanda-wimp}
J.~Ahrens et al., Phys.\ Rev.\ {\bfseries D66} (2002) 032006.

\bibitem{superk-wimp}
A. Habig {\em et al.} {\it Proceedings of  the XVII
International Cosmic Ray Conference (ICRC)}, Hamburg, Germany, 2001, p. 1558. Also hep-ex/0106024;
S~Desai, talk at Identificaiton of Dark Matter, 2002 (idm2002), York, England.

\bibitem{icecube-wimp}
J.~Edsj\"o, internal Amanda/IceCube report, 2000.

\end{thebibliography}
\end{document}